\documentclass[12pt]{article}
\pdfoutput=1

\usepackage[utf8]{inputenc}

\usepackage{draft}
\usepackage{putex}
 
\usepackage[all]{xy}

\usepackage{stackrel,amssymb}

\usepackage{graphicx}
\usepackage{caption}
\usepackage{amsmath,bm}
\usepackage{array}
\usepackage{subcaption}
\usepackage{epstopdf}
\usepackage{enumerate}
\usepackage{cite}
\usepackage{tensor}
\usepackage{slashed}
\usepackage[utf8]{inputenc}
\usepackage{rotating}
\usepackage{bigfoot}
\usepackage[
      colorlinks=true,
      linkcolor=blue,
      urlcolor=blue,
      filecolor=black,
      citecolor=red,
      ]{hyperref}

 \usepackage{adjustbox}
\usepackage{multirow}
\usepackage{tikz-cd}

\def\XXint#1#2#3{{\setbox0=\hbox{$#1{#2#3}{\int}$ }
		\vcenter{\hbox{$#2#3$ }}\kern-.6\wd0}}

\numberwithin{equation}{section}

\def\<{\langle}
\def\>{\rangle}

\def\pa{\partial}

\def\ep{\epsilon}

\def\cI{{\cal I}}

\newcommand{\leftrarrows}{\mathrel{\raise.75ex\hbox{\oalign{%
				$\scriptstyle\leftarrow$\cr
				\vrule width0pt height.5ex$\hfil\scriptstyle\relbar$\cr}}}}
\newcommand{\lrightarrows}{\mathrel{\raise.75ex\hbox{\oalign{%
				$\scriptstyle\relbar$\hfil\cr
				$\scriptstyle\vrule width0pt height.5ex\smash\rightarrow$\cr}}}}
\newcommand{\Rrelbar}{\mathrel{\raise.75ex\hbox{\oalign{%
				$\scriptstyle\relbar$\cr
				\vrule width0pt height.5ex$\scriptstyle\relbar$}}}}

\makeatletter
\def\leftrightarrowsfill@{\arrowfill@\leftrarrows\Rrelbar\lrightarrows}
\newcommand{\xleftrightarrows}[2][]{\ext@arrow 3399\leftrightarrowsfill@{#1}{#2}}
\makeatother

\begin{document}

\preprint{}

	\institution{CMSA}{Center of Mathematical Sciences and Applications, Harvard University, Cambridge, MA 02138, USA}
	\institution{HU}{Jefferson Physical Laboratory, Harvard University,
		Cambridge, MA 02138, USA}

\title{
Surface Defect, Anomalies and $b$-Extremization 
}

\authors{Yifan Wang\worksat{\CMSA,\HU}}

\abstract{
Quantum field theories (QFT) in the presence of defects exhibit new types of anomalies which play an important role in constraining the defect dynamics and defect renormalization group (RG) flows. Here we study surface defects and their anomalies in conformal field theories (CFT) of general spacetime dimensions. When the defect is conformal, it is characterized by a conformal $b$-anomaly analogous to the $c$-anomaly of 2d CFTs. The $b$-theorem states that $b$ must monotonically decrease under defect RG flows and was proven by coupling to a spurious defect dilaton. We revisit the proof by deriving explicitly the dilaton effective action for defect RG flow in the free scalar theory. For conformal surface defects preserving $\cN=(0,2)$ supersymmetry, we prove a universal relation between the $b$-anomaly and the 't Hooft anomaly for the $U(1)_r$ symmetry. We also establish the $b$-extremization principle that identifies the superconformal $U(1)_r$ symmetry from $\cN=(0,2)$ preserving RG flows. Together they provide a powerful tool to extract the $b$-anomaly of strongly coupled surface defects.
To illustrate our method, we determine the $b$-anomalies for  a number of surface defects in 3d, 4d and 6d SCFTs. We also comment on manifestations of these defect conformal and 't Hooft anomalies in defect correlation functions.
}
\date{}

\maketitle

\tableofcontents

\pagebreak

\section{Introduction and setup} 

A fundamental problem in the study of quantum field theories (QFT) is to uncover nonperturbative constraints on renormalization group (RG) flows. A powerful tool to tackle this problem comes from anomalies for the relevant symmetries. In particular, the 't Hooft anomaly for a symmetry $G$ must be constant along $G$-symmetric RG flows, leading to nontrivial constraints on the low energy dynamics, known as the \textit{anomaly matching condition} \cite{tHooft:1979rat}. The 't Hooft anomalies can often be extracted without knowing the details of the QFT dynamics, thanks to the robustness under small deformations of the theory, providing a quick and valuable diagnostic on  candidate RG flows and infra-red (IR) phase diagrams from a given ultra-violet (UV) description.  

Much subtler are the anomalies for accidental symmetries that only emerge at ends of the RG flow, the epitome of which is the conformal symmetry at RG fixed points and the associated conformal anomalies in even spacetime dimensions. The hallmark of conformal symmetry is the existence of a locally conserved, symmetric and traceless stress tensor $T_{\m\n}$, which gives rise to a conformal field theory (CFT). Importantly the traceless condition can be violated by contact terms, or equivalently by certain curvature terms when the theory is placed on curved manifolds, that solve the associated Wess-Zumino consistency conditions \cite{Wess:1971yu},
\ie
\la T^\m_\m \ra= -(-1)^{d/2} a E_d +\sum_i c_i W_i\,,
\label{trgen}
\fe
where $E_d$ is the Euler class in $d$ dimensions normalized such that $\int_M E_d=\chi(M)$ with $\chi(S^d)=2$ for an even dimensional sphere, and $W_i$ are Weyl invariants of the curvature.
Equivalently, the trace anomaly contributes to the anomalous variation of the path integral under the Weyl transformation $g\to g e^{2\sigma}$ where $g$ is the metric on the spacetime manifold $\cM$,
\ie
\D_\sigma \log Z[g]=\int_\cM d^d x \sqrt{g} \,\sigma  \left( -(-1)^{d/2} a E_d +\sum_i c_i W_i \right)\,.
\label{weylvarZ}
\fe
The coefficients $a$ and $c_i$ are the conformal anomaly coefficients. Unlike the usual 't Hooft anomalies, the conformal anomalies are only defined at the end points of RG flows and do not match since conformal symmetry is broken along the flow. Instead  they give rise to powerful inequality constraints on RG flows. Indeed it has long been expected that a combination of the conformal anomalies measures degrees of freedom in the CFT and thus should decrease under RG flow which is intuitively an irreversible coarse-graining procedure that produces IR dynamics from UV descriptions. For $d=2$, the answer is affirmative as shown in \cite{Zamolodchikov:1986gt} where the only conformal anomaly coefficient is $a=c_{2d}$. In fact there exists a locally defined $c$-function on the space of 2d QFTs which coincides the conformal anomalies at the fixed points and monotonically decreases along all RG flows, proving the so-called $c$-theorem which establishes the irreversibility of RG flows for $d=2$ \cite{Zamolodchikov:1986gt}. Shortly after it was conjectured in \cite{Cardy:1988cwa} that the $a$-anomaly in \eqref{trgen} is the appropriate generalization of $c_{2d}$ to even $d\geq 4$ and the corresponding $a$-theorem was postulated. It was only until recently the $a$-theorem was proven in $d=4$ \cite{Komargodski:2011vj,Komargodski:2011xv,Casini:2017vbe}, and some progress has been made in extending the proof of \cite{Komargodski:2011vj,Komargodski:2011xv} to $d=6$ with success in special cases \cite{Elvang:2012st,Cordova:2015vwa,Cordova:2015fha}. Notably, by introducing a dilaton to compensate for the broken conformal symmetry along the RG flow, the authors of \cite{Komargodski:2011vj,Komargodski:2011xv} recasted the problem into one of the 't Hooft anomaly matching type, and deduced the $a$-theorem based on unitarity constraints on the dilaton effective action.

The studies of QFTs are greatly enriched by the incorporation of defects, which are ubiquitous in nature and can come from boundaries, interfaces or higher-codimension impurities in quantum systems, and are just starting to be systematically explored. 
Local deformations on the defect worldvolume give rise to defect RG flows that produce an array of new defect critical phenomena.
A natural task is to identify the new defect anomalies that constrain these defect RG flows and the defect phase diagram. The usual 't Hooft anomalies have straightforward generalizations for defect symmetries.  The trace of the stress tensor  can also receive new anomaly contributions localized on the defect worldvolume at the critical point. 

Here we focus on the case of two dimensional defect, which can be a boundary or interface in $d=3$ or more generally a surface defect in higher dimensions. In this case, the most general defect 't Hooft anomalies for continuous symmetries take the following form. Let us denote the surface defect by $\cD$ and its worldvolume submanifold by $\Sigma$. As we vary the full partition function $Z_\cD[A,e]$ in the presence of a background gauge connection $A$ for the defect global symmetry $G$ as well as the vielbein  and spin connection $(e,\omega)$ by gauge transformations $\D_\lambda A= d\lambda$ and $\D_\theta \omega= d\theta+[\omega, \theta]$, we have
\ie
\D_{\theta,\lambda} \log Z_\cD[A,e] \supset i\int_\Sigma  \left( \kappa_1\tr (\lambda F) + \kappa_2\theta R  + \kappa_3\tr(\lambda)R+ \kappa_4\theta \tr(F)   \right)\,,
\label{agen}
 \fe
where $F$ and $R$ are the corresponding curvature two-forms. In the above, we have only included the anomalous variations intrinsic to the defect, which does not include the bulk anomalies that may be present without the defect. The defect 't Hooft anomalies are captured by the coefficients $\kappa_i$ (the last two in \eqref{agen} correspond to mixed anomalies which are possible if $G$ contains abelian factors). 
Like their bulk counterparts, these defect 't Hooft anomalies can often be obtained by deforming the coupled system to a weakly interacting description thanks to the topological nature of these anomalies.

At the critical point of the bulk-defect coupled system, we have a conformal surface defect in the ambient CFT, and the full theory is sometimes referred to as defect CFT  (DCFT). A flat defect $\cD$ preserves the $\mf{so}(2,2)$ conformal subalgebra and is thus expected to share many features of 2d CFTs. However a crucial difference between the conformal defect and a local CFT is the generic absence of a locally conserved stress tensor on the defect. More generally, symmetries on $\cD$ can, but do not necessarily, lead to locally conserved currents. This is of course consistent due to the mild non-locality on the defect worldvolume. Thus apart from the familiar anomalies for 2d CFTs,  we have the following anomalous contributions to the Ward identities of bulk conserved currents. For an abelian global symmetry current $J^f_\m$ in the bulk, such defect anomalies take the form
\ie
\la \nabla_\m J_f^\m \ra_\cD \supset {1\over 4\pi}\D(\Sigma)   \star_\Sigma(i k_{ff} F_f+\sum_{I\neq r} i k_{fI} F_I  +k_{fg} R)\,,
\label{aJ}
\fe
where $F_r$ and $F_I$ are curvatures for background abelian gauge fields. The coefficient $k_{ff}$ characterizes the pure anomaly associated to the symmetry of $J^r_\m$, while $k_{fI}$ and $k_{fg}$ are mixed anomalies with other abelian flavor and Lorentz symmetries.
As for the stress tensor $T_{\m\n}$, there is potentially a defect gravitational anomaly\footnote{Here and below we only include anomalous contributions that purely depend on the metric, as the mixed anomalies are already captured by \eqref{aJ}.
}
\ie
\la \nabla_\m T^{\m a } \ra_\cD \supset  {k_g\over 24\pi} \D(\Sigma) \ep^{ab}\nabla_b R_\Sigma  \,,
\fe
and a defect conformal anomaly whose general form is given by 
\cite{Graham:1999pm,Henningson:1999xi,Schwimmer:2008yh} 
\ie
\la T^\m_\m \ra_\cD\supset 
{1\over 24\pi} \D(\Sigma)\left(
 b R_\Sigma + d_1  (K^i_{ab} K_i^{ab}-{1\over 2}K^i K_i)  - d_2 W_{abcd} h^{ac}h^{bd}
\right)
\label{trgend}
\,.
\fe
Here we have split the spacetime coordinates  into directions tangential and normal to $\Sigma$ as $x^\m=(x^a,y^i)$.  We denote the induced metric on $\Sigma$ by $h_{ab}$,  the Ricci curvature scalar on $\Sigma$ by $R_\Sigma$, the extrinsic curvature by $K_{ab}^i$,  and the pull back of the bulk Weyl curvature to $\Sigma$ by $W_{abcd}$. Correspondingly the Weyl variation of the path integral $\D_\sigma \log Z_\cD [g] $, compared to \eqref{weylvarZ}, receives defect localized contributions
\ie
 \cA^{\rm Weyl}_\cD\equiv  {1\over 24\pi } \int_\Sigma d^2 x \sqrt{h} \sigma  
\left(
 b R_\Sigma + d_1  (K^i_{ab} K_i^{ab}-{1\over 2}K^i K_i)  - d_2 W_{abcd} h^{ac}h^{bd}
\right)\,.
\label{Weylav}
\fe

In this paper, we will be mostly interested in the conformal (gravitational) anomalies $b$ and $k_g$. To make connection to known results about 2d CFTs, we define  
\ie
c_L \equiv b-{k_g\over 2},\quad c_R\equiv b+{k_g\over 2}\,,
\fe
which coincide with the chiral central charges of a local 2d CFT in the degenerate case where the bulk CFT is trivial and the stress tensor is only supported on the defect. Under defect RG flows, the gravitational anomaly $k_g$ stays constant. On the other hand the $b$-anomaly plays the role of the $c$-function for surface defects and the $b$-theorem states that $b$ decreases monotonically under defect RG flows. The $b$-theorem is proven in \cite{Jensen:2015swa} by employing a defect version of the dilaton effective action method of \cite{Komargodski:2011vj,Komargodski:2011xv}. In Section~\ref{sec:dilaton}, we revisit the defect $b$-theorem and present an explicit derivation of the dilaton effective action for the boundary RG flow of a free scalar field in $d=3$, which also provides 
a check on the boundary $b$-anomalies obtained from heat kernel methods \cite{Vassilevich:2003xt}. However a generic surface defect does not have Lagrangian descriptions and it has remained challenging to extract the $b$-anomalies for interacting defects, which we will overcome in this work. As we explain in Section~\ref{sec:sda}, the conformal anomalies of a large class of interacting surface defects can be derived in terms of their 't Hooft anomalies which are much easier to compute. This is made possible by the presence of $\cN=(0,2)$ supersymmetry on the defect worldvolume $\Sigma$. The method relies on an defect version of the $c$-extremization principle \cite{Benini:2012cz,Benini:2013cda} which we prove here. 
We emphasize that our results are completely non-perturbative and do not rely on any Lagrangian descriptions.
In Section~\ref{sec:examples}, we apply our methods to selected examples of superconformal surface defects in various spacetime dimensions. It is straightforward to reproduce recent results of $b$ anomalies based on  large $N$ holography \cite{Gentle:2015jma,Rodgers:2018mvq,Jensen:2018rxu,Estes:2018tnu} and supersymmetric localization when the defect has enhanced supersymmetries \cite{Chalabi:2020iie}. We also describe examples for which the previous methods do not apply, such as the boundary $b$-anomalies of 3d $\cN=2$ Ising and SQED SCFTs. We end with a short summary and discuss future directions in Section~\ref{sec:discuss}.

\section{Defect $b$-theorem and dilaton effective action}
\label{sec:dilaton}

We start by reviewing the proof of the defect $b$-theorem \cite{Jensen:2015swa} using the spurious dilaton  \cite{Komargodski:2011vj,Komargodski:2011xv}. Let us consider a defect RG flow from a UV conformal defect $\cD$. We formally restore the defect conformal symmetry by introducing a non-dynamical  dilaton field $\tau$ localized on the defect worldvolume $\Sigma$, which shifts $\tau \to \tau +\sigma$  under local Weyl rescaling $g_{\m\n} \to e^{2 \sigma} g_{\m\n}$. Consequently the conformal (Weyl) anomaly remains constant for the full system along the RG flow. In the IR, the effective action takes the following form,
\ie
S_{\rm eff}=S_{\cD_{\rm IR}} +S_{\tau} + \dots 
\fe
where $S_{\cD_{\rm IR}}$ abstractly describes the IR DCFT, $S_\tau$ is the dilaton effective action, and we have omitted coupling between the IR DCFT and $\tau$ which is suppressed in the derivative expansion (higher than second order) \cite{Jensen:2015swa}. Anomaly matching then demands
\ie
\D_\sigma \log  Z_{\cD }  =\D_\sigma \log Z_{\cD_{\rm IR}}-\D_\sigma S_\tau\,,
\fe
which fixes the form of $S_\tau$ up to terms that are diffeomorphism and Weyl invariant \cite{Komargodski:2011xv,Jensen:2015swa}. In particular for a flat defect in flat space, we simply have 
\ie
S_\tau= {\Delta b\over 24\pi}\int_\Sigma d^2 x \, \pa_a \tau \pa^a \tau   +  \int_\Sigma d^2 x \,\Lambda^2 e^{-2\tau}\,.
\label{staugen}
\fe 
where $\Delta b\equiv b_{\rm UV} -b_{\rm IR}$. On the other hand, since by construction $\tau$ couples to the defect local operator $\cT$ in $T^\m_\m=\D(\Sigma)\cT$ for a flat defect, we have \cite{Komargodski:2011xv,Jensen:2015swa}
 \ie
 \Delta b=3\pi \int_\Sigma d^2x |x|^2 \la \cT(x)\cT(0)\ra 
 \fe
and unitarity (reflection positivity) requires $\Delta b>0$.

To illustrate better the defect dilaton effective action which plays a crucial role in the proof, we find it educational to consider the simplest example of a nontrivial defect RG flow, namely that between the Neumann and Dirichlet boundary conditions of a free scalar field in 3d. We will explicitly derive the dilaton effective action in this case, and find agreement with known results about the defect $b$-anomalies from heat kernel methods at the defect fixed points. 

We start with the Euclidean free scalar action  
\ie
S={1\over 2}\int_{y\geq 0} d^3 x \pa_\m \Phi \pa^\m \Phi \,,
\fe
on $\mR_+^3$ with Neumann boundary condition
\ie
\left. \pa_3 \Phi \right|_\Sigma=0\,.
\fe
We then turn on a relevant perturbation in the form of a boundary mass term
\ie
\Delta S = -{1\over 2} \int d^2 x \, m\Phi^2\,.
\fe
This deforms the boundary condition to the mixed type
\ie
\left.\pa_3 \Phi  -m \Phi \right |_\Sigma=0\,,
\fe
which interpolates between the Neumann boundary condition at $m=0$ and Dirichlet boundary condition $\Phi|_\Sigma=0$ as $m \to \infty$.

Now we introduce the defect dilaton $\tau$ to restore the defect conformal symmetry,
\ie
S_{\rm tot}={1\over 2}\int_{y\geq 0} d^3 x \pa_\m \Phi \pa^\m \Phi -{1\over 2} \int_\Sigma d^2 x \, m  e^\tau\Phi^2
\fe
and the goal is to derive the effective coupling for $\tau$ from integrating out $\Phi$. Since the action is quadratic, this can be easily accomplished by Wick contractions once we have the two-point function (propagator),
\ie
 G(\vec x,\vec x',y,y') \equiv \la \Phi(\vec x,y) \Phi(\vec x',y')\ra\,,
\fe
with the general boundary condition
\ie
\left. \pa_y \Phi(\vec x,y)-m \Phi(\vec x,y)\right|_{y\to 0^+}=0\,.
\fe
The solution is  given by a $m$-dependent linear combination of the free scalar propagator (in the absence of boundary) between $x$ and $x'$, and that between  its mirror image $\bar x \equiv (x,-y)$ and $x'$. Here we find it convenient to perform a Fourier transform in the $\vec x$ coordinates, and the desired two-point function is the following linear combination
\ie
\hat G(p,-p,y,y')=
{e^{-|p||y-y'|}\over 2|p|} +{|p|-m\over  |p|+m } {e^{-|p|(y+y')}\over 2|p|}
\fe
up to a prefactor that impose momentum conservation in the boundary directions. It satisfies
\ie
\left. (\pa_y-m) \hat G(p,-p,y,y')\right|_{y\to 0^+}=0
\fe
as desired. The effective action for $\tau$ can be determined by computing the boundary correlators of $\Phi^2$. For this purpose, we simply need
\ie
\hat  G(p,-p,0,0)=
{1\over |p|+m}\,,
\fe
and the  one-loop Feynman diagram in a large $m$ expansion
\ie
\int d^2x e^{i \vec p \cdot \vec  x}\la  \Phi^2(\vec x,0) \Phi^2(0,0)\ra_c = 2\int {d^2 k\over (2\pi)^2}{1\over (|k|+m)(|k+p|+m)} = \A_1 +\A_2 {p^2\over  m^2}+\cO \left({p^4\over  m^4} \right)\,.
\fe
In particular, the two-derivative term in the dilaton effective action is 
\ie
-{\A_2\over 8} \int d^2x  \, \pa_a \tau \pa^a \tau \,.
\fe
Explicit computation (see Appendix~\ref{appendix:bdyf}) gives 
 \ie
 \A_2=-{1\over 24 \pi}\,.
 \fe
 Compared to \eqref{staugen}, we conclude for the boundary RG flow of a free 3d scalar field
 \ie
 \Delta b={1\over 8}
 \fe
This is indeed consistent with the fixed point values of the $b$-anomalies computed from heat kernel methods \cite{Jensen:2015swa}
\ie
b_{\rm Dir}=-{1\over 16},\quad b_{\rm Neu}={1\over 16}\,.
\fe
In comparison, a real 2d scalar on $\Sigma$ contributes $b=1$.

\section{Defect anomalies and supersymmetry}
\label{sec:sda}

Let us now consider unitary superconformal surface defects preserving $\cN=(0,2)$ supersymmetry. According to the classification of \cite{nw2}, such defects can exist  in 3d $\cN\geq 2$, 4d $\cN\geq 1$ and 6d $(\cN\geq 1,0)$ SCFTs. Familiar examples include boundaries and interfaces in 3d, as well as surface operators in 4d and 6d SCFTs, sometimes preserving an enhanced superconformal symmetry (see Section~\ref{sec:examples} for detailed examples). In all cases, the defect preserves a $U(1)_r$ symmetry of the bulk SCFT, which is identified with the $R$-symmetry of the $\cN=(0,2)$ superconformal algebra 
\ie
\mf{sl}(2,\mR)\oplus \mf{osp}(2|2,\mR)\,.
\label{dsca}
\fe
 In this section, we will derive the following universal relation between  the defect $b$-anomaly and the 't Hooft anomaly of $U(1)_r$  given by $k$,
\ie
b+{k_g\over 2}=3k\,,
\fe
or equivalently
\ie
c_L=3k-k_g,\quad c_R=3k \,.
\label{ckrel}
\fe
Note that for standalone 2d $\cN=(0,2)$ SCFTs, these relations are automatically satisfied. In those case the anomaly coefficients appear in the OPE of 2d local stress tensor and $R$-symmetry currents, and \eqref{ckrel} is a simple consequence of the SUSY Ward identities. Here we will show that the same relation persists for defect anomaly coefficients, even when such 2d local conservation laws are no longer present.

\subsection{'t Hooft anomalies and inflow}
We start by reviewing the 't Hooft anomalies of a general surface defect $\cD$ (not necessarily conformal). For the anomalies associated to diffeomorphism and abelian symmetries which couple to background gauge fields $A^I$, the corresponding anomaly polynomial is given by,
\ie
\cI_4=-{k_g\over 24} p_1(T)+{1\over 2} k_{IJ} c_1(F^I) c_1(F^J)= {k_g\over 192 \pi^2} \tr  R\wedge R -  {k_{IJ}\over 8\pi^2} F^I\wedge F^J
\fe
which determines the anomalous variation of $\log Z_\cD$ by the usual descent procedure. Equivalently, the defect anomalies are produced by the inflow from the CS action
\ie
S_{\rm CS}=  {i\over 4\pi} \int_{\cM} \left(   8\pi k_g{\rm CS}_g -k_{IJ}A^I d F^J \right)
\label{bCS}
\fe
to its boundary $\Sigma=\pa \cM$. Here ${\rm CS}_g$ is the gravitational Chern-Simons term satisfying
\ie
  d{\rm CS}_g
={1\over 192\pi}\tr R\wedge R\,.
  \fe
Note that ${\rm CS}_g$ is defined up to a shift by an exact 3-form, which corresponds to the Bardeen-Zumino counter-term on $\Sigma$ which shifts between anomalies for diffeomorphism and Lorentz symmetries \cite{Bardeen:1984pm}.

To ease subsequent comparisons to known results in the literature, our normalization is such that for a complex right-moving Weyl fermion $\chi$ on $\Sigma$ that carry charge $q$ under the $U(1)$ symmetry,
\ie
\cI_4(\chi)=  {1\over 192 \pi^2} \tr  R\wedge R - {q^2\over 8\pi^2} F\wedge F\,.
\fe

\subsection{Conformal $b$-anomaly from 't Hooft anomalies and SUSY}  
For the superconformal defect  $\cD$, it can happen that the bulk stress tensor $T_{\m\n}$ is related to some symmetry current $J_\m$ by supercharges $\cQ$  preserved by the defect (i.e. $[\cQ, \cD]=0$). In such a scenario, the trace anomaly $T^\m_\m$ and 't Hooft anomaly $\pa_\m J^\m$ are related by SUSY Ward identities, as is well-known  for bulk anomalies in even dimensions. For $\cN=(0,2)$ superconformal surface defects, the relevant symmetry current $J_\m$ is the one that generates the defect superconformal symmetry $U(1)_r$. It is in general a linear combination that involves a bulk $R$-symmetry current $J^R_\mu$ and possibly the current $x_{[i}T_{j]\m}$ that generates rotation transverse to the defect (see \cite{nw2} for explicit expressions). The current $J^\m$ is preserved by the defect
\ie
\la \pa_\m J^\m \ra_\cD=0\,,
\fe
in the absence of background couplings. Nevertheless, as reviewed previously, the $U(1)_r$ symmetry can have a 't Hooft anomaly localized on the defect,
\ie
\la \pa_\m J^\m \ra_\cD= {k\over 4\pi}iF\D^2(\Sigma) \,.
\fe
When coupled to background gauge field $A$ on the entire spacetime manifold $\cM$, it contributes to an anomalous variation under a gauge transformation $\lambda$ on the defect worldvolume $\Sigma$,
\ie
 \D_{\lambda} \log Z_\cD= {k\over 4 \pi}  i
 \int_\Sigma    \,\lambda F\,,
 \label{Rvar}
 \fe
 which is local on $\Sigma$.
For the superconformal defect, the anomalous variations with respect to other symmetry transformations are governed by the SUSY completion. 

We first focus on the case without gravitational anomalies, namely $k_g=0$ for the defect $\cD$.\footnote{Strictly speaking we also assume that the right-moving $U(1)_r$ symmetry can be extended to a non-anomalous non-holomorphic $U(1)_A$ symmetry which is gauged by the background $\cN=(0,2)$ supergravity \cite{Gomis:2015yaa}. However this can be easily achieved by tensoring the defect $\cD$ with an auxiliary $\cN=(2,2)$ anomaly-free SCFT (e.g. the supersymmetric $T^2$ sigma model) of $c_L=c_R=3$ and $k_L=k_R=1$. Denoting the generator of the left-moving R-symmetry in this free $\cN=(2,2)$ SCFT by $r_L$ and the right-moving R-symmetry of the full system by $r$, we identify the combination $\sqrt{k+1}r_L-r$ as the generator for the non-anomalous $U(1)_A$ symmetry that is gauged in the $\cN=(0,2)$ supergravity. The rest of the argument is unaffected.}
The $\cN=(0,2)$ completion of \eqref{Rvar} is fixed by SUSY and diffeomorphism invariance \cite{Gomis:2015yaa},
\ie
 \D_{\Omega} \log Z_\cD=i{k\over 4 \pi} \left(
 \int d^2 x  d\theta^+ \,\D\Omega \cR_- 
 -c.c.\right)\,.
 \label{SWeylav}
 \fe
 Here $\theta_+$ is the Grassmann coordinate in the $(0,2)$ superspace, $\cR_-$ is the curvature superfield\footnote{Gauge fields are anti-Hermitian in this paper and compared to \cite{Gomis:2015yaa} $F_{\rm there}= i F_{\rm here}$.}
   \ie
  \cR_-=-{i\over 4} \theta^+(R_\Sigma\sqrt{g}+2 F) + {\rm fermions}
  \fe
 and $\D \Omega$ a chiral superfield whose bottom component $\sigma+i\lambda$ packages together the Weyl and $U(1)_r$ transformation parameters. Comparing \eqref{SWeylav} to \eqref{Weylav}, we conclude 
 \ie
 b=c_L=c_R=3k\,.
 \label{bknga}
 \fe

\subsection{Gravitational anomalies and Chern-Simons counterterms}
In the previous section, we focused on surface defect $\cD$ with vanishing gravitational anomaly $k_g=0$. 
Here we will lift this restriction and arrive at the general result \eqref{ckrel}.

For defects with $k_g \in {\mZ\over 2}$, we can cancel the gravitational anomaly by introducing decoupled free fields that respect the $\cN=(0,2)$ supersymmetry. The argument from the previous section then proceeds for the total system without gravitational anomaly and the result \eqref{ckrel} follows after subtracting off the free field contributions. 
More generally one can cancel the gravitational anomaly by a supersymmetric version of the anomaly inflow mechanism. The CS term \eqref{bCS} has a supersymmetric generalization \cite{Rocek:1985bk,Achucarro:1987vz,Achucarro:1989gm,Closset:2012ru,Closset:2012vg} that preserves 3d $\cN=2$ superconformal symmetry on a closed manifold,
  \ie
S_{\rm sCS}= 2i \kappa \int_{\cM_3}  \left({\rm CS}_g- {1\over 48\pi} B\wedge dB
  +  {\rm fermions} 
  \right)
  \label{sCS}
  \fe
where $B$ is the 3d $U(1)_R$ gauge field.\footnote{Note that gauge fields in the paper are anti-Hermitian. This leads to a relative minus sign in \eqref{sCS} compared to the expression in \cite{Closset:2012ru,Closset:2012vg}.}
If $\cM_3$ has boundary $\pa \cM_3=\Sigma$, with suitable boundary terms,\footnote{Although we will not derive these boundary terms here, let us make a few comments. The Chern-Simons term \eqref{sCS} being superconformal naturally plays a role in $\cN=2$ supergravity on AdS$_3$. In particular they account for the gravitational anomalies of boundary $\cN=(0,2)$ 2d SCFTs. Bosonic parts of the boundary terms were proposed \cite{Kraus:2005zm,Kraus:2006wn} but the full supersymmetric completion is unknown to the author's knowledge.  
It would be desirable to derive the full boundary term from $\cN=2$ supergravity, perhaps by extending the work of
\cite{Belyaev:2007bg} for $\cN=1$ supergravity.} we expect to preserve the half-BPS $\cN=(0,2)$ supersymmetry and $B$  restricted to $\Sigma$ is the identified with the $U(1)_r$ background gauge field $A$.

Taking $\kappa=-k_g$, the inflow from \eqref{sCS} shifts both the gravitational and the $U(1)_r$ anomalies. Since the Chern-Simons term \eqref{sCS} does not contribute to Weyl anomalies,\footnote{A quick way to see this is to perform a constant Weyl transformation and \eqref{sCS} is clearly invariant. Under a general Weyl rescaling, the gravitational Chern-Simons action $\int_{\cM_3} {\rm CS}_g$ changes by a boundary term  on $\Sigma$ which corresponds to mixed Lorentz-Weyl anomalies (or mixed diffeomorphism-Weyl anomalies)  \cite{Chamseddine:1992ry}.
}
the shifts are 
\ie
   k \to k-{k_g\over 6},\quad c_R \to c_R-{k_g\over 2},\quad c_L \to c_L+{k_g\over 2}\,. 
  \fe
Since the total system with the auxiliary $\cM_3$ and Chern-Simons term \eqref{sCS} is free of gravitational anomaly and respects $\cN=(0,2)$ supersymmetry, we conclude
\ie
c_R-{k_g\over 2}=3\left(k-{k_g\over 6}\right)
\fe
from \eqref{bknga} in the previous section, and the result \eqref{ckrel} follows.

 \subsection{Defect RG flow and $b$-extremization}
\label{sec:cex} 

In the previous sections, we have shown that at the conformal fixed point of the defect field theory, the conformal anomalies of an $\cN=(0,2)$ surface defect are completely determined by its $U(1)_r$ and gravitational anomalies as in \eqref{ckrel}. Since 't Hooft anomalies are preserved along symmetric RG flows, it is natural to ask whether such relations can be used to determine the $b$-anomaly of the IR conformal defect from an effective description of the defect at some immediate scale along the flow. As we will show, this is accomplished by the $b$-extremization principle, which is a simple extension of the $c$-extremization principle that applies to  standalone 2d $\cN=(0,2)$ theories.

Let us consider a defect RG flow that preserves $\cN=(0,2)$ Poincar\'e supersymmetry, a $U(1)_{\hat r}$ R-symmetry under which the supercharges have charges $\hat r=\pm 1$  and a number of abelian global symmetries $q_I$ that commute with the supercharges. In the absence of accidental symmetries along the defect RG flow, the IR superconformal $U(1)_r$ symmetry is generally a linear combination
\ie
r=\hat r +\sum_I t^I_{\star} q_I\,.
\fe
We define the trial $b$-anomaly as a quadratic polynomial in the mixing parameters $t^I$,
\ie
b_{\rm trial}(t^I)=3(k_{\hat r\hat r}+2 \sum_I t^I k_{\hat rI} +\sum_{I,J} t^I t^J k_{IJ})-{k_g\over 2}\,,
\label{btr}
\fe
whose coefficients $k_{\hat r \hat r},k_{\hat r I},k_{IJ}$ characterize the 't Hooft anomalies among the $U(1)_{\hat r}$ and abelian symmetries $q_I$.\footnote{In practice, we will work with the scheme such that $k_{IJ}$ is symmetric.} Its value at $t^I=t^I_\star$ yields the actual $b$-anomaly according to \eqref{ckrel},
\ie
b = b_{\rm trial}(t^I_{\star})\,.
\fe
The $b$-extremization principle states that 
\ie
\left. {\pa  b_{\rm trial} (t^I)\over \pa t^I} \right|_{t^I=t^I_\star}=0
\label{bext}
\fe
which determines the value of $t^I_\star$ uniquely and thus the IR $b$-anomaly follows. 

The proof of \eqref{bext} is similar to that in the original work on $c$-extremization \cite{Benini:2012cz,Benini:2013cda} with a new ingredient that the current $J_\m$ for $U(1)_r$ R-symmetry is not locally conserved on the defect worldvolume $\Sigma$. Let's suppose we are at the IR $\cN=(0,2)$ superconformal fixed point of a defect RG flow. The $U(1)_r$ current is  generally a linear combination\footnote{Here we remind the readers that we split the bulk spacetime coordinates as $x^\m=(x^a,y^i)=(\vec x,\vec y)$ into directions tangential and transverse to the defect worldvolume $\Sigma$.
For surface defects, we will use $x^a$ and $(z,\bar z)$ interchangeably.}
\ie
&J_\m(x)=   J^\cB_\m(x) + \D_\m^{\bar z} J^{\cD}_{\bar z} (\vec x) \D^{d-2}(\vec y)
\fe
of some bulk current $J_\m^\cB$ satisfying
\ie
 \pa^\m J_\m^\cB(x) =0\,,
\fe
and defect right-moving current $J^\cD_{\bar z}$ satisfying
\ie
\pa_{z} J^\cD_{\bar z}(\vec x)=0\,.
\fe
When no such defect conserved currents exist on $\cD$, the $U(1)_r$ current $J_\m$ is \textit{irreducible}, namely its $\cN=(0,2)$ descendants generate an irreducible defect stress-tensor superconformal multiplet.\footnote{For related works on surface defect stress-tensor multiplets (not necessarily conformal), see for example \cite{Drukker:2017dgn,Brunner:2019qyf}.} An obvious example of \textit{reducible} $J_\m$ is when the DCFT contains a decoupled $\cN=(0,2)$ SCFT with its own stress-tensor multiplet whose primary is a locally conserved 2d R-current.  

The condition \eqref{bext} is equivalent to the vanishing of the mixed $U(1)_r$ anomaly $k_{rI}=0$ for the superconformal defect. Since
\ie
\la \pa_\m J^\m (\vec x,\vec y) j^I_a(\vec x') \ra ={k_{r I}\over 4\pi} \ep_{ab} \pa^b \D^2(\vec x-\vec x')\D^{d-2} (\vec y)\,,
\fe
it suffices to show that
\ie
\la J_\m(\vec x,\vec y) j^I_a(\vec x')\ra=0\,,
\fe
for any defect conserved current $j^I_a(\vec x)$.

We start with the case where $J_\m$ is irreducible. The two-point functions of bulk and defect primary vectors are highly constrained by the residual conformal symmetry\cite{Billo:2016cpy,Herzog:2020bqw}. For vector primary operators, the relevant two-point function is fixed up to two constants,\footnote{We include a parity-odd structure which was not considered in \cite{Billo:2016cpy,Herzog:2020bqw}.}
\ie
\la J_\m(\vec x,y) j_a^I(\vec x') \ra={c_{\rm even} I_{\m a}+c_{\rm odd} I_{\m}{}^b \ep_{ab} \over s^2 |\vec y|^{d-2}}  \,,
\fe
where $s_\m\equiv (\vec x-\vec x',\vec y)$ and
\ie
I_{\m a}\equiv \D_{\m a}-{2 s_\m s_a \over |s|^2}
\fe
is the defect version of the familiar symmetric tensor coming from the Jacobian for inversion transformation. Imposing the current conservation conditions, we conclude
\ie
c_{\rm even}=c_{\rm odd}=0\,.
\fe

Now if $J_\m$ is reducible, we can write
\ie
&J_\m(x)= J^\cB_\m(x) + \D_\m^{\bar z} J^{\cD}_{\bar z} (\vec x) \D^{d-2}(\vec y)
\fe
where $J^\cB_\m$ and $J^{\cD}_{\bar z}$ are separately conserved, and $J^\cB_\m$ is irreducible.  By the previous arguments, it then suffices to show that
\ie
\la J^{\cD}_{\bar z} (\vec x)j_a^I(\vec x')\ra =0\,,
\label{dJJ}
\fe
which follows from the explanations in \cite{Benini:2012cz,Benini:2013cda} which we repeat below (in a somewhat different way) for completeness. 

From $\cN=(0,2)$ superconformal symmetry, the defect $U(1)_r$ current $J^{\cD}_{\bar z}$ (a reducible component of the full $U(1)_r$ current) resides in a supercurrent multiplet \cite{Dumitrescu:2011iu},
\ie
\cJ^\cD=J^\cD_{\bar z} -i \theta^+ G^\cD_{\bar z +} -i \bar\theta^+ \bar G^\cD_{\bar z +}- \theta^+ \bar  \theta^+ T^\cD_{\bar z\bar z}
\fe
where $G_{\bar z+}^\cD,\bar G_{\bar z+}^\cD$ are right-moving defect supercurrents and $T^\cD_{\bar z\bar z}$ is the defect stress-tensor. Note that all of these operators are conserved locally on the defect. 
If $j_a^I$ is left-moving, \eqref{dJJ} is obvious by conformal symmetry. For right-moving currents $j_a^I$, we denote them as $j_{\bar z}^I$ and consider the three-point function
\ie
\la j_{\bar z}^I (\bar z_1) G^\cD_{\bar z +} (\bar z_2)   \bar G^\cD_{\bar z +}(0)  \ra \,.
\fe
The above three-point function vanishes from inspecting the $j_{\bar z}^I(z_1)  G^\cD_{\bar z +}(z_2)$ OPE since by assumption the $\cN=(0,2)$ supercharges and thus the supercurrent $G^\cD_{\bar z +}$ are uncharged under $j_{\bar z}^I$. On the other hand, from the ${1\over \bar z^2 }J^\cD_{\bar z}(0)$ term in the $G^\cD_{\bar z +} (\bar z)   \bar G^\cD_{\bar z +}(0)$ OPE we conclude $\la J^{\cD}_{\bar z} j_{\bar z}^I\ra=0$ as desired. This completes the proof of the $b$-extremization principle.

Before we end this section, let us comment on a caveat in the arguments above that was also present for $c$-extremization in \cite{Benini:2012cz,Benini:2013cda}. Here we have assumed that the currents $j^I_a$ are conformal primaries and whose right-moving and left-moving components are separately conserved, which follow from unitarity and the existence of a normalizable conformally invariant vacuum for the DCFT. In particular this excludes the possibility of a non-compact complex scalar $\phi$ (which completes to a $\cN=(0,2)$ chiral multiplet) on the defect. In that case, the right-moving non-primary current $\pa_{\bar z} \phi$ can mix with the $U(1)_r$ current in the presence of a nontrivial background charge on the surface defect (i.e. due to the coupling $q\int_\Sigma d^2x \sqrt{h}R_\Sigma {\rm Re}\phi $ and its SUSY completion). The $b$-extremization principle continues to hold in this case with the understanding that the resulting $b$-anomaly is really the effective anomaly defined as
\ie
b_{\rm eff}=b-24\bar h_{\rm min}\,,
\fe
where $\bar h_{\rm min}$ is the minimal $\bar L_0$ eigenvalue among the defect local operators, closely related to the effective conformal charge $c_{\rm eff}$ defined in \cite{Kutasov:1990sv}. In particular the proper generalization of Zamolodchikov's $c$-theorem \cite{Zamolodchikov:1986gt} to non-compact CFTs uses $c_{\rm eff}$  \cite{Hori:2001ax,Adams:2001sv,Harvey:2001wm}. Similarly we expect $b_{\rm eff}$ to be the monotonic quantity under defect RG flows when the defect field theory is non-compact.  

A further related subtlety when there is no normalizable conformally invariant vacuum in the DCFT is the appearance of non-holomorphic conserved currents whose left and right moving components are not separately conserved. This can also be illustrated in the context of a free non-compact $\cN=(0,2)$ chiral multiplet \cite{Benini:2013cda}. As explained there,  the $c$-extremization principle \eqref{bext} does not hold when $j^I_a$ is non-holomorphic and irreducible, and the same problem arises for the $b$-extremization of defect conformal anomaly. Nevertheless in practice one can try to isolate such non-holomorphic currents and extremize $b_{\rm trial}$ among the rest of the currents to determine the conformal $b$-anomaly.

\section{Examples of $\cN=(0,2)$ surface defects and anomalies}
\label{sec:examples}
There has been promising recent progress in understanding the conformal anomalies of $\cN=(0,2)$ superconformal surface defects, namely $b$ and $d_{1,2}$ in \eqref{trgend}. From Wess-Zumino consistency conditions, it is obvious that the $b$-anomaly cannot depend on marginal couplings on the surface defect.\footnote{This is a common feature of the \textit{$a$-type} conformal anomalies, whose Weyl variations are total derivatives.} 
Furthermore it was shown in \cite{Bianchi:2019umv} that with $\cN=(0,2)$ supersymmetry $b$ cannot depend on bulk marginal couplings either. 

For $\cN=(0,2)$ surface defects in 4d SCFTs, it was proven in \cite{Bianchi:2019sxz} using SUSY Ward identities that $d_1=d_2$. Given the relation between $d_{1}$ and the displacement operator two-point function $c_D$, and between $d_2$ and the stress tensor one-point function $h$ \cite{Lewkowycz:2014jia,Bianchi:2015liz}, it was explained in \cite{Bianchi:2019sxz} how to determine the $d_{1,2}$ anomalies from knowledge of the chiral algebra   underlying the 4d $\cN\geq 2$ SCFT \cite{Beem:2013sza,Cordova:2017mhb}. The $b$-anomaly is however more elusive, and expected to enter in the two-point function of bulk stress tensor in the presence of the surface defect. Alternatively, it can be accessed from the $S^4$ partition function of the SCFT decorated with the surface defect on $S^2$ by inspecting the   logarithmic dependence of the free energy on the sphere radius. When a localization formula  \cite{Nekrasov:2003rj,Pestun:2007rz} for such a setup exists (i.e. for $\cN=(2,2)$ surface defects in 4d $\cN=2$ SCFTs with gauge theory descriptions \cite{Drukker:2010jp,Kanno:2011fw,Gomis:2014eya,Gorsky:2017hro}), this was implemented in \cite{Chalabi:2020iie} to identify the $b$-anomaly. We will see how to recover these results easily using \eqref{ckrel} which does not rely on the localization formulae. 

Conformal anomalies of $\cN=(4,4)$ superconformal surface defects in 6d $(2,0)$ SCFTs have been studied in \cite{Berenstein:1998ij,Henningson:1999xi,Henningson:1999iw,Gustavsson:2003hn,Gustavsson:2004gj,Gentle:2015jma,Rodgers:2018mvq,Jensen:2018rxu,Mezei:2018url,Estes:2018tnu,Chalabi:2020iie,Drukker:2020dcz,Drukker:2020swu,Drukker:2020atp}. In particular, the relation $d_1=d_2$ was shown to persist for the half-BPS surface defects in 6d $(2,0)$ SCFTs \cite{Drukker:2020atp}, and their values are determined by the defect superconformal index \cite{Bullimore:2014upa} as explained in \cite{Chalabi:2020iie}. However the $b$-anomalies have only be obtained in the free theory \cite{Drukker:2020dcz} and in the large $N$ limit from holography \cite{Gentle:2015jma,Rodgers:2018mvq,Jensen:2018rxu,Estes:2018tnu}. Here we will give exact answers for $b$ for general half-BPS surface defects in 6d $(2,0)$ SCFTs of arbitrary ADE types and the generalization to $\cN=(0,4)$ surface defects in 6d $(1,0)$ SCFTs is straightforward.

Finally little is known about the conformal anomalies of boundaries in 3d CFTs apart from some attempts from holography \cite{Estes:2018tnu}, thus we will be most pedagogical with this case in the following, to illustrate our method by determining the $b$-anomalies for several types of simple $\cN=(0,2)$ superconformal boundaries. We leave the more sophisticated examples that involve non-abelian Chern-Simons-matter bulk SCFTs to future investigation.

\subsection{Boundaries in 3d  SCFTs}
In $d=3$, the relevant $\cN=(0,2)$ superconformal defect is either a boundary or an interface of some 3d $\cN\geq 2$ SCFT. 
Since the interface between two theories $\cT_1$ and $\cT_2$ is related by the folding trick to the boundary in the tensor product theory $\cT_1 \times  \bar \cT_2$ (the second factor involves an orientation-reversal), we will focus on boundary defects here without loss of generality. 
For a 3d $\cN=2$ bulk SCFT, the $U(1)_r$ symmetry of the half-BPS DCFT is identified with the $U(1)_R$ symmetry of the bulk SCFT. The identification for $\cN>2$ SCFTs is similar (involves the $U(1)_R$ of an $\cN=2$ subalgebra) and the detailed mapping can be found in \cite{nw2}. Examples of $\cN=(0,2)$ boundaries and interfaces can be found in
 \cite{Berman:2009kj,Berman:2009xd,Dimofte:2012pd,Gadde:2013wq,Gadde:2013sca,Okazaki:2013kaa,Yoshida:2014ssa,Bullimore:2016nji,Dimofte:2017tpi}. We will follow the conventions of \cite{Dimofte:2017tpi} here.

\subsubsection{Free theories}
\label{sec:3dfree}
We start by considering the boundary conditions of the free 3d $\cN=2$ SCFT made out of a single chiral multiplet $\Phi_{\rm 3d}$ which consists of a complex scalar $\phi$ and a Dirac fermion $\psi_\pm $ whose $U(1)_r$ charges are $r={1\over 2}$ and $r=-{1\over 2}$ respectively. The theory lives  on the half space $\mR^{2,1}_+$ defined by $y\geq 0$ with boundary $\Sigma =\mR^{1,1}$ at $y=0$.

To study $\cN=(0,2)$ preserving boundary conditions, it is convenient to decompose the bulk degrees of freedom into representations of the $\cN=(0,2)$ subalgebra.
Here the 3d chiral multiplet $\Phi_{\rm 3d}$  decomposes into a $\cN=(0,2)$ chiral multiplet
\ie
\Phi=\phi+\theta^+\psi_+-i\theta^+\bar \theta^+\pa_{\bar z}\phi\,,
\fe
and a Fermi multiplet
\ie
\Psi=\bar\psi_- +\theta^+ f-i \theta^+\bar\theta^+ \pa_{\bar z}\psi_-\,,
\fe
where $f$ is an auxiliary field that satisfies  $f =\pa_y \bar \phi$ on-shell \cite{Dimofte:2017tpi}. Here $\pm$ are indices for right and left moving spinors as before. Both sub-multiplets have $U(1)_r$ charge $r={1\over 2}$.

The basic supersymmetric boundary conditions of $\Phi_{3d}$ involve setting either $\Phi$ or $\Psi$ to zero on $\Sigma$, corresponding to supersymmetric Dirichlet and Neumann boundary conditions \cite{Berman:2009kj,Okazaki:2013kaa,Dimofte:2017tpi}:
\ie
\cB_D[\Phi_{3d}]:~\Phi|_\Sigma=0 \to  \phi |_\Sigma =\psi_+|_\Sigma =0\,,
\fe
and
\ie
\cB_N[\Phi_{3d}]:~\Psi|_\Sigma=0 \to \pa_y \phi |_\Sigma =\psi_-|_\Sigma =0\,.
\fe
The 3d fermions contribute nontrivial boundary 't Hooft anomalies and can be worked out by considering mass deformations \cite{Dimofte:2017tpi}. The results are summarized in Table~\ref{tab:bta}.

\begin{table}[!htb]
\centering
    \renewcommand{\arraystretch}{1.8}
        \begin{tabular}{|c|c|c|}
            \hline
          & Fields         & Anomaly $\cI_4$    \\ \hline
            \multirow{2}{*}{$2d$} &   $\chi_-$  & $-{1\over 2}q^2 c_1(F)^2+{1\over 24} p_1(T)$ 
            \\\cline{2-3}
                          &  $\chi_+$ & ${1\over 2} q^2c_1(F)^2-{1\over 24} p_1(T)$     \\\hline
             \multirow{3}{*}{$3d$} &    $\left.\psi_+\right|_\Sigma=0$   & $-{1\over 4} q^2c_1(F)^2+{1\over 48} p_1(T)$ 
            \\\cline{2-3}
                          &   $\left.\psi_-\right|_\Sigma =0$   & ${1\over 4}q^2 c_1(F)^2-{1\over 48} p_1(T)$   
                          \\\cline{2-3}
                          &  $\left.{k\over 4\pi} AdA \right|_{y\geq 0}$     & $-{k\over 2} c_1(F)^2$  \\\hline    
        \end{tabular}
    \caption{The 't Hooft anomalies for 2d Weyl fermions $\chi_\pm$, 3d Dirac fermions $\psi_\pm$ with different boundary conditions, and a classical 3d CS action on $\mR^{2,1}_+$. Here the fermions carry charge $q$ under the vector $U(1)$ symmetry for which $A$ is the background gauge connection.}
    \label{tab:bta}
\end{table}

For either boundary conditions, since there are no extra currents on the boundary, the $b$-anomalies (or equivalently $c_L,c_R$) are easily determined by \eqref{ckrel} from the $U(1)_r$ and gravitational anomalies,
\ie
\cB_D[\Phi_{3d}]:&~k=-{1\over 8},~k_g= -{1\over 2}
,\quad 
\cB_N[\Phi_{3d}]:k={1\over 8},~ k_g=  {1\over 2}\,.
\fe
The results are tabulated in Table~\ref{tab:ndclcr}. In particular we find perfect agreement with known results about $b$-anomalies of free scalar and fermion on $\mR^{2,1}$ \cite{Jensen:2015swa,Fursaev:2016inw}. We emphasize that unlike the conformal anomalies of standalone CFTs, the defect conformal anomalies do not need to be positive.\footnote{It would be interesting to see if there is a universal lower bound on the $b$-anomaly of conformal boundary (surface) defects. See some relevant discussions in \cite{Jensen:2018rxu}.}

\begin{table}[!htb]
\centering
      \renewcommand{\arraystretch}{1.2}
        \begin{tabular}{|c|c|c|c|}
            \hline
          &   Fields        &  $c_L$ &  $c_R$  \\ \hline
            \multirow{5}{*}{$2d$}  &   $\chi_-$  & 1  & 0
            \\\cline{2-4} &   $\chi_+$  &  0 & 1
             \\\cline{2-4} &   $\varphi$  &  2 & 2
            \\\cline{2-4}&   Fermi  &  1 & 0
            \\\cline{2-4}
                         &   chiral  &  2 & 3   \\\hline
             \multirow{6}{*}{$3d$} &    $\left.\psi_-\right|_\Sigma=0$   & $-{1\over 4}$  & ${1\over 4}$  
            \\\cline{2-4}
                          &   $\left.\psi_+\right|_\Sigma =0$   & $1\over 4$  & $-{1\over 4}$  
            \\\cline{2-4}
                          &   $\left.\pa_y \phi \right|_\Sigma =0$   & $1\over 8$  & $ {1\over 8}$ 
            \\\cline{2-4}
                          &   $\left. \phi \right|_\Sigma =0$   & $-{1\over 8}$  & $-{1\over 8}$ 
            \\\cline{2-4}
                          &   $\cB_N[\Phi_{3d}]$   & $-{1\over 8}$  & $ {3\over 8}$ 
            \\\cline{2-4}
                          &  $\cB_D[\Phi_{3d}]$     & ${1\over 8}$  & $-{3\over 8}$               \\\hline    
        \end{tabular}
    \caption{The top entries give the conformal anomalies for 2d Weyl fermions $\chi_\pm$, complex scalar $\varphi$ and $\cN=(0,2)$ Fermi and chiral multiplets. The bottom entries give the boundary conformal anomalies of 3d Dirac fermion $\psi$, complex scalar $\phi$ with basic boundary conditions and their $\cN=(0,2)$ supersymmetric completions.}
    \label{tab:ndclcr}
\end{table}

The $\cN=(0,2)$ Dirichlet and Neumann boundary conditions are related by supersymmetric defect RG flows \cite{Dimofte:2017tpi}. This is achieved by coupling the boundary conditions with an extra free 2d chiral multiplet $C$ or Fermi multiplet $\Gamma$, and turning on  superpotential deformations (known as \textit{flip} from \cite{Dimofte:2012pd}) as follows,
\ie
&\cB_D [\Phi_{3d}] \oplus C  ~{\rm with}~\int_\Sigma d^2x d\theta^+  C \Psi ~\longrightarrow~ \cB_N [\Phi_{3d}]
\\
&\cB_N [\Phi_{3d}]  \oplus \Gamma  ~{\rm with}~\int_\Sigma d^2x d\theta^+ \Phi \Gamma   ~~\longrightarrow~  \cB_D [\Phi_{3d}]\,.
\fe
From Table~\ref{tab:ndclcr}, we see clearly the defect $b$-theorem is obeyed for these simple RG flows. 

Given two 3d chiral multiplets $\Phi_{3d}$ and $\Phi'_{3d}$ with Dirichlet and Neumann boundary conditions respectively, there is a superpotential deformation that couples the two multiplets together at the boundary
\ie
&\cB_D [\Phi_{3d}] \oplus \cB_N [\Phi'_{3d}]   ~{\rm with}~\int_\Sigma d^2x d\theta^+    \Psi \Phi'
\label{tcg}
\fe
which is exactly marginal, and amounts to an $\cN=(0,2)$ preserving rotation of the original boundary conditions for $(\Phi_{3d},\Phi'_{3d})$. Via the unfolding trick, we have $\Phi_{3d}$ and $\Phi'_{3d}$ on $\mR^{2,1}_+$ and $\mR^{2,1}_-$ respectively joined at $\Sigma$. Then the superpotential in \eqref{tcg} implements the identification between $(\Phi,\Psi)$ and $(\Phi',\Psi')$ along the interface at $\Sigma$, so that the total system is simply $\Phi_{3d}$ on $\mR^{2,1}$ with a transparent interface.\footnote{To see this, one uses the following boundary variations of the chiral multiplet action on the two half spaces \cite{Dimofte:2017tpi}
\ie
\D S[\cB_D [\Phi_{3d}]]=\int_\Sigma  d^2x d\theta^+ \D \Psi \Phi+c.c.,\quad 
\D S[\cB_N [\Phi'_{3d}]]=\int_\Sigma  d^2x d\theta^+ \Psi' \D\Phi'+c.c.\,.
\fe
} This explains why $\cB_D [\Phi_{3d}]$ and $\cB_D [\Phi_{3d}]$ have opposite 't Hooft and defect conformal anomalies as in Table~\ref{tab:bta} and \ref{tab:ndclcr}.

\subsubsection{$\cN=2$ Ising SCFT}
Let us now consider the 3d $\cN=2$ Ising SCFT which is defined by a single 3d chiral multiplet $\Phi_{3d}$ with bulk superpotential $W=\Phi_{3d}^3$. Here the $U(1)_r$ charge of $\Phi_{3d}$ is $r={2\over 3}$. 

The simplest superconformal boundary condition is the Dirichlet boundary $\cB_D[\Phi_{3d}]$. Since there are no extra abelian symmetries on the boundary, the boundary conformal anomalies simply follow from the $U(1)_r$ and gravitational 't Hooft anomalies, contributed by the Dirac fermion $\psi_\pm $ in the chiral multiplet,
\ie
k=-{1\over 18},\quad k_g=-{1\over 2}\,.
\fe
Thus from \eqref{ckrel} we conclude
\ie
c_L={1\over 3},\quad c_R=-{1\over 6}\,.
\fe
There are more interesting boundary conditions coming from coupling the Ising SCFT with Dirichlet boundary condition to nontrivial $\cN=(0,2)$ SCFTs on the boundary but we will leave that to future work.

\subsubsection{$\cN=2$ SQED and mirror symmetry}
When the 3d SCFT has gauge theory descriptions, there is a plethora of interesting supersymmetric boundary conditions that are expected to flow to superconformal boundaries in the IR \cite{Dimofte:2017tpi}. We will study them more systematically in a future publication and focus on the simplest example here, namely the $\cN=2$ $U(1)$ SQED with one chiral multiplet of charge 1.  Since this theory is mirror dual of a free chiral multiplet, we will also be able to make connections to boundary conditions of the free SCFT.

The 3d $\cN=2$ vector multiplet $V_{3d}$ has components $(A_\m,\lambda_\pm,\sigma,D)$ where $D$ is an auxiliary field. Here we consider the Dirichlet boundary condition  \cite{Dimofte:2017tpi}
\ie
\cB_D[V_{3d}]:~\left. A_a \right|_\Sigma = \left.\lambda_- \right|_\Sigma = \left.D \right|_\Sigma=0\,,
\fe
together with $\cB_D[\Phi_{3d}]$ for the charged chiral multiplet. We will denote the full boundary condition as $\cB_D[{\rm SQED}]$.

The 3d theory has a $U(1)_{\hat{R}}$ symmetry under which the fermions in the vector and chiral multiplets have charges
\ie
\hat{R}[\lambda_\pm]= 1,\quad \hat{R}[\psi_\pm ]= -1\,.
\fe
In addition, there is a topological global symmetry $U(1)_T$ which maps to the flavor symmetry that rotates the free chiral multiplet in the mirror dual. Correspondingly, the free chiral multiplet itself is described by a BPS monopole operator and its super-partners in the SQED. The 3d superconformal R-symmetry is the combination
\ie
R=\hat R+{1\over 2}T\,,
\fe
such that the BPS monopole has $R={1\over 2}$ saturating the 3d unitarity bound.

A novelty of the Dirichlet boundary condition for gauge field $\cB_D[V_{3d}]$ is the presence of additional 2d global symmetry $U(1)_G$ that comes from the gauge symmetry in the bulk. Consequently, the $U(1)_R$ symmetry can mix with $U(1)_G$ in the presence of the Dirichlet boundary, and the superconformal $U(1)_r$ symmetry generator will be a linear combination of the generators $R$ and $G$. Below we will see how $b$-extremization fixes this linear combination and thus the conformal defect anomalies in this case. 

We start by recalling the boundary 't Hooft anomalies for $U(1)_R$ and $U(1)_G$ for the SQED  given in \cite{Dimofte:2017tpi}
\ie
\cI_4= \cI_4^{\rm UV}
+{1\over 4}c_1(F_{\hat R})^2-{1\over 4}(c_1(F_G)-c_1(F_{\hat R}))^2-{k_g\over 24} p_1(T)\,.
\label{sqedta}
\fe
where
\ie
\cI_4^{\rm UV}=-{1\over 4}(c_1(F_G)-c_1(F_{\hat R})^2
- c_1(F_T)c_1(F_G)+{1\over 4}c_1(F_{\hat R})^2
-{1\over 4}(c_1(F_T)-c_1(F_{\hat R})^2
\label{SQEDUVcs}
\fe
comes from UV Chern-Simons couplings required for the duality between SQED and free chiral multiplet in the absence of a boundary.\footnote{ Here we have included a shift (the last term in \eqref{SQEDUVcs}) compared to the expression in \cite{Dimofte:2017tpi}, to match with our regularization scheme for the free chiral multiplet in Section~\ref{sec:3dfree}.} This includes a $U(1)_{1\over 2}$ Chern-Simons coupling for the gauge field to ensure gauge invariance.
The second term of \eqref{sqedta} comes from boundary anomalies of the gaugino $\lambda_\pm$, and the third term is due to the fermions $\psi_\pm$ in the chiral multiplet.

The last term in \eqref{sqedta} encodes the boundary gravitational anomaly which can be obtained by giving large negative mass $m<0$ to both $\lambda_\pm $ and $\psi_\pm$. Integrating out the fermions, we end up with a $U(1)_0$ scalar QED which is free of gravitational anomalies. But we also need to remember that the boundary condition $\left. \lambda_- \right |_\Sigma=\left. \psi_+ \right|_\Sigma =0$ supports an edge mode of $\psi_-$ for $m<0$, which contributes to the boundary gravitational anomaly
\ie
k_g=-{1\over 2}\,.
\fe
Equivalently we can consider large positive mass $m>0$ for the bulk fermions, which leads to a $U(1)_1$ Maxwell-Chern-Simons theory coupled to scalars, which contributes $k_g=-1$ since $U(1)_1$ is equivalent to the gravitational Chern-Simons action $e^{-2i\int {\rm CS}_g}$ \cite{Seiberg:2016gmd}. Now for $m>0$, the edge mode consistent with the boundary condition comes from $\lambda_+$, thus we recover the same total boundary gravitational anomaly as above. 

To summarize, the full boundary 't Hooft anomalies for $\cB_D[{\rm SQED}]$ are given by
\ie
\cI_4= -{1\over 2}c_1(F_G)^2- c_1(F_G)(c_1(F_T)-c_1(F_{\hat R})-{1\over 4}
(c_1(F_T)-c_1(F_{\hat R}))^2+{1\over 48} p_1(T)\,.
\label{sqedtafull}
\fe
Let us now write down the trial $b$-anomaly \eqref{btr} using these 't Hooft anomalies in \eqref{sqedtafull}, with the candidate $U(1)_r$ symmetry generated by $r=R+t G =\hat R+{1\over 2}T + t G$ with parameter $t$,
\ie
b_{\rm trial}(t)=3k_{rr}+{1\over 4}=3(-t^2+t-{1\over 8})+{1\over 4}\,.
\fe
Extremizing with respect to $t$, we find that the superconformal $U(1)_r$ symmetry for the Dirichlet boundary condition of SQED is
\ie
r= \hat R+ {1\over 2}T+ {1\over 2} G\,,
\fe
and the defect conformal anomalies are given by
\ie
\cB_D[{\rm SQED}]:\quad c_L={7\over 8},\quad c_R={3\over 8}\,.
\fe
In \cite{Dimofte:2017tpi} it was proposed that under mirror symmetry, $\cB_D[{\rm SQED}]$ is dual to $\cB_N[\Phi_{3d}]$ for the free 3d chiral multiplet with an extra free Fermi multiplet on the boundary. Recall the defect conformal anomalies of a free 3d chiral multiplet,
\ie
\cB_N[\Phi_{3d}]:\quad c_L=-{1\over 8},\quad c_R={3\over 8}\,.
\fe
We see the difference is precisely saturated by that of a 2d Fermi multiplet (see Table~\ref{tab:ndclcr}).

\subsection{Surface defects in 4d SCFTs}
 There is a rich zoo of surface defects in 4d SCFTs preserving $\cN=(0,2)$ or a further enhanced superconformal symmetry, such as the half-BPS surface defects in $\cN=1$ SCFTs \cite{Koh:2009cj,Drukker:2017dgn,Razamat:2018zel} and those in $\cN=2,4$ SCFTs \cite{Gukov:2006jk,Gomis:2007fi,Drukker:2008wr,Koh:2008kt,Alday:2009fs,Gaiotto:2009fs}. 
 
One of the most well-studied example is the half-BPS Gukov-Witten (GW) surface operator in the $\cN=4$ super-Yang-Mills (SYM) \cite{Gukov:2006jk}, which is defined by a codimension-two singularity in the SYM fields. The resulting defect enjoys small $\cN=(4,4)$ superconformal symmetry $\mf{psu}(1,1|2)\times \mf{psu}(1,1|2)$ in the IR which contains \eqref{dsca} as an $\cN=(0,2)$ subalgebra, thus we can determine the defect conformal anomalies from the 't Hooft anomalies following our general arguments in the previous section. For this purpose, it is convenient to use an alternative UV description of the same surface defect, as a 2d-4d system, that involves coupling the 4d gauge theory with gauge group $G$ on $\cM$ to an auxiliary 2d field theory on $\Sigma$ \cite{Gukov:2006jk}.\footnote{More precisely, the $\cN=(4,4)$ superconformal defect is described by the Higgs branch of the auxiliary 2d theory \cite{Gadde:2013dda}. See \cite{Witten:1997yu} for discussions on related subtleties in $\cN=(4,4)$ RG flows.} Typically the coupling is through gauging a $G$ flavor symmetry on the defect. For $SU(N)$ SYM, the GW surface defects are labelled by a partition $N=\sum_{i=1}^n k_i$.\footnote{For general gauge group $G$, the GW surface defects are labelled by the Levi subgroups $\mathbb{L}$ of $G$. Here for $G=SU(N)$, we have $\mathbb{L}=S[U(k_1)\times \dots \times U(k_n)]$ corresponding to the partition $N=\sum_i k_i$.} The corresponding auxiliary 2d theory is described by a 2d $\cN=(4,4)$ linear quiver gauge theory, with gauge group (except for the bold node which is a flavor symmetry) $U(p_1)\times \dots U(p_{n-1}) \times \bm{SU(p_n)}$ and bifundamental hypermultiplets between each pair of consecutive nodes \cite{Gadde:2013dda}. Here the rank of the gauge nodes are
\ie
p_j=\sum_{i=1}^j k_i,\quad p_n=N\,.
\fe

The 't Hooft anomalies of the surface defect follow immediately from the field content of the auxiliary 2d theory. 
In particular there is no gravitational anomaly in this $\cN=(4,4)$ gauge theory, and the anomaly of the superconformal $U(1)_r$ is simply related to the (right-moving)  superconformal $SU(2)_R$  anomaly by
\ie
k = 2k_R\,.
\label{kkR}
\fe
Recall the 2d gauge theory has $SU(2)_L\times SU(2)_R\times SU(2)_I$ R-symmetry in the UV and the superconformal R-symmetry on the Higgs branch is identified with $SU(2)_L\times SU(2)_R$ \cite{Witten:1997yu}. The chiral fermions in the theory, $\lambda_\pm$ from the vector multiplet and $\psi_\pm$ from the hypermultiplet, transform under $SU(2)_L\times SU(2)_R\times SU(2)_I \times SU(2)_F$ as
\ie
&(\lambda_+,\lambda_-):~(2,1,2,1)_+\oplus (1,2,2,1)_-,
\\
&(\psi_+,\psi_-):~(1,2,1,2)_+\oplus (2,1,1,2)_-,
\fe
where we have introduced $SU(2)_F$ to keep track of the global symmetry of a free hypermultiplet.
Thus the right-moving $SU(2)_R$ superconformal R-symmetry  receives anomaly from $\lambda_-$ and $\psi_+$ in the quiver gauge theory,
\ie
 k_R=\sum_{i=1}^{n-1} (p_i p_{i+1}- p_i^2)
={1\over 2} \left( N^2-\sum_{i=1}^n k_i^2 \right)\,.
 \fe
 Consequently from \eqref{kkR} we have determined the conformal anomalies of a general Gukov-Witten surface defect in $SU(N)$ SYM
 \ie
 b=c_L=c_R=6k_R=3\left( N^2-\sum_{i=1}^n k_i^2 \right)\,.
 \fe
 
GW surface defects in SYM with general gauge group $G$ are labelled by Levi subgroups $\mathbb{L}\subset G$. We will denote them by $\cD_{\mathbb{L}}[G]$. The auxiliary 2d theory can be described by a $\cN=(4,4)$ non-linear sigma model with hyperK\"ahler target space $T^*(G/\mathbb{L})$ \cite{Gukov:2006jk}, thus
 \ie
 b=c_L=c_R=6k_R=3\left( \dim (G)-\dim (\mathbb{L}) \right)\,.
 \fe
 
 We observe that for these surface defects, the defect conformal anomalies are nothing but the usual conformal anomalies of the auxiliary 2d theory (in the IR conformal limit). This is not a coincidence. In the description of surface defects by 2d-4d systems, 
 the 2d theory is coupled to the 4d SCFT by bulk gauge fields, thus  they decouple in the weak coupling limit $g_{\rm YM}\to 0$. Since $b$-anomalies of surface defects with $\cN=(0,2)$ supersymmetry do not depend on bulk marginal couplings \cite{Bianchi:2019umv}, they must coincide with the conformal anomalies of the 2d theory viewed as a standalone SCFT (in the IR). This gives a quick way to determine the defect conformal anomalies for a large class of surface defects in 4d $\cN=2$ conformal gauge theories, and easily reproduces the localization results found in \cite{Chalabi:2020iie}.
 
\subsection{Surface defects in 6d SCFTs}
Despite the non-Lagrangian nature of the 6d SCFTs, their string/M/F-theory constructions suggest that they host interesting surface defects. For example, in M-theory, such surface operators arise from two-dimensional M2-M5 intersections \cite{Lunin:2007ab}, possibly in the presence of transverse singularities \cite{DelZotto:2014hpa} and/or Horava-Witten walls \cite{Horava:1996ma}. They define half-BPS $\cN=(0,4)$ or $\cN=(4,4)$ surface defects in 6d $\cN=(1,0)$ and $\cN=(2,0)$ SCFTs respectively. The preserved superconformal algebra is  $\mf{sl}(2,\mR)\oplus \mf{osp}(4^*|2)$ for the $\cN=(0,4)$ defect  and $\mf{osp}(4^*|2)\oplus \mf{osp}(4^*|2)$ for the $\cN=(4,4)$ case.\footnote{Note that this is a different superconformal algebra compared to the small $\cN=(4,4)$ algebra preserved by a half-BPS surface defect in the 4d $\cN=4$ SYM. This difference is crucial in determining the defect conformal anomalies from its 't Hooft anomalies.} In either case, the $\cN=(0,2)$ superconformal algebra \eqref{dsca} is a subalgebra and thus our results from the previous sections apply. We will need the following relation between the $U(1)_r$ generator of the $\cN=(0,2)$ subalgebra and R-symmetry generators in $\mf{osp}(4^*|2)$,
\ie
r= 2(R- 2I)\,.
\label{rRF}
\fe
Here $R$ and $I$ are the Cartan generators of the $\mf{su}(2)_R\times \mf{su}(2)_I$ R-symmetry of the right-moving $\mf{osp}(4^*|2)$, normalized with half-integral eigenvalues.\footnote{The factors of 2 in \eqref{rRF} are important. See Appendix B.3 of \cite{Agmon:2020pde} for explicit expressions.} Then if we know the 't Hooft anomalies for the surface defect, we can use \eqref{ckrel} to determine the conformal anomalies. Note that because of the enhanced R-symmetry (and supersymmetry), we don't expect mixing with global symmetries. 

For illustration, let us work out the anomalies for a class of $\cN=(4,4)$ surface defects in the 6d $(2,0)$ theory labelled by an ADE Lie algebra $\mf{g}$. The surface defect is in general characterized by a weight vector $\lambda \in \Lambda_{\rm w}(\mf{g})$, and we will refer to it by $\cD_\lambda[\mf{g}]$. Upon compactification on an $S^1$ longitudinal to the defect, the bulk SCFT is described by 5d $\cN=2$ SYM with gauge algebra $\mf{g}$ and the surface defect $\cD_\lambda[\mf{g}]$ corresponds to a half-BPS Wilson loop in the representation with highest weight $\lambda$ \cite{Bullimore:2014upa,Mezei:2018url}. 

The 't Hooft anomalies of the surface defect $\cD_\lambda[\mf{g}]$ can be deduced by moving onto the tensor branch of the 6d SCFT since the R-symmetries and Lorentz symmetry are preserved. 

On a generic point of the tensor branch in a general 6d $(1,0)$ SCFT, the effective action takes the following schematic form  
\ie
S_{\rm TB} = 2\pi \int \eta^{ij} \left( 
{1\over 2} dB_i\wedge \star dB_j +B_i\wedge I_j
\right)+\dots 
\fe
where $B_i$ denotes the self-dual 2-form field for each tensor multiplet, $\eta^{ij}$ is a symmetric, positive-definite and integral charge matrix,
and $I_i\wedge B_j$ where $I_i$ is a 4-form made of characteristic classes in background gauge fields and geometry is the Green-Schwarz term \cite{GREEN1985327} that plays an important role in the matching of 6d 't Hooft anomalies on the tensor branch \cite{Ohmori:2014kda,Intriligator:2014eaa}. 
On the tensor branch, the $\cN=(0,4)$ surface defect is expected to be described by a BPS self-dual string  of charge $Q_i$ under the 2-form fields $B_i$.\footnote{See for example \cite{Haghighat:2013gba,Haghighat:2013tka,Kim:2014dza,Haghighat:2014vxa,Gadde:2015tra} for works on the self-dual strings in various 6d $(1,0)$ theories.} From anomaly inflow \cite{Kim:2016foj,Shimizu:2016lbw}, the corresponding anomaly polynomial is determined
\ie
\cI_4={1\over 2} \eta^{ij}Q_i Q_j (c_2(F_L)-c_2(F_R))+  \eta^{ij}Q_i I_j\,.
\label{gensta}
\fe
 
Now let us come back to the particular case of  $\cD_\lambda[\mf{g}]$ defects in 6d $(2,0)$ SCFTs. To apply \eqref{gensta}, we note that here $\eta^{ij}$ is the Cartan matrix of $\mf{g}$, the string charge satisfies $\lambda= \sum_i Q_i \A_i$ where $\A_i$ are the simple roots of $\mf{g}$, and the Green-Schwarz term takes a simple form with  \cite{Ohmori:2014kda,Shimizu:2016lbw}
\ie
I_i=\rho_i (c_2(F_I)-c_2(F_F))
\fe
where $\rho=\sum_i \rho_i \A_i$ is the Weyl vector of $\mf{g}$. Therefore we can write
\ie
\cI_4[\cD_\lambda[\mf{g}]]={1\over 2} (\lambda,\lambda) (c_2(F_L)-c_2(F_R))+  (\lambda,\rho) (c_2(F_I)-c_2(F_F))\,.
\label{I420}
\fe
Note the absence of gravitational anomalies $k_g=0$ in this case. 
The $\cN=(4,4)$ surface defect (and the corresponding string on the tensor branch) has $SO(4)_L \times SO(4)_R$ R-symmetry, and $SO(4)_L=SU(2)_L\times SU(2)_F$ wheres  $SO(4)_R=SU(2)_R\times SU(2)_I$ \cite{Shimizu:2016lbw}. From \eqref{rRF} and \eqref{I420}, we can read off the 't Hooft anomaly for the $U(1)_r$ symmetry of the $\cN=(0,2)$ subalgebra,
\ie
k=(\lambda,\lambda)+8 (\lambda,\rho)\,.
\fe
Consequently we obtain the conformal anomalies for the surface defect $\cD_\lambda[\mf{g}]$ from \eqref{ckrel},
\ie
b=c_L=c_R=3(\lambda,\lambda)+24 (\lambda,\rho)\,.
\fe
This agrees with the results of \cite{Estes:2018tnu} for $\mf{g}=\mf{su}(N)$ obtained from holographic entanglement entropy in the presence of the defect. 

\section{Discussion}
\label{sec:discuss} 

In this paper, we have studied the defect analog of the 2d conformal anomalies, namely the $b$-anomaly, for unitary conformal surface defects in CFTs. We revisited the defect $b$-theorem and provided an explicit example of the dilaton effective action for defect RG flows in the free scalar theory. Such defect dilaton  effective action played an important role in the proof of \cite{Jensen:2015swa}. We also investigated 't Hooft anomalies that arise in the presence of a surface defect. For defects with $\cN=(0,2)$ superconformal symmetry, we derived a universal relation between the $b$-anomaly and the 't Hooft anomaly of $U(1)_r$ symmetry. Since the latter is much more robust against deformations, this provides a shortcut to determine the $b$-anomalies of strongly coupled conformal defects using weak coupling results that are typically available after deformations. A potential subtlety arises when trying to identify the superconformal defect $U(1)_r$ symmetry away from the conformal fixed point. This is settled by the $b$-extremization principle that governs defect RG flows with $\cN=(0,2)$ supersymmetry  which we have proved in this work. The $b$-extremization picks out the superconformal $U(1)_r$ symmetry among the symmetries preserved by an RG flow, whose 't Hooft anomaly then determines the conformal $b$-anomaly, in analogue to the well-known $c$-extremization principle for standalone 2d $\cN=(0,2)$ theories \cite{Benini:2012cz,Benini:2013cda}. To illustrate our method, we then set off to determine the $b$-anomalies for a number of surface defects in 3d, 4d and 6d SCFTs. We now discuss some future directions below.

\subsubsection*{Holographic dual of $b$-extremization and $b$-anomalies}
The $c$-extremization principle of 2d $\cN=(0,2)$ SCFTs \cite{Benini:2012cz,Benini:2013cda} has an elegant geometric dual in the context of AdS/CFT \cite{Couzens:2018wnk,Gauntlett:2019roi,vanBeest:2020vlv}, in terms of extremizing certain functionals of off-shell geometries in string/M/F-theory (in close analogy to the holographic dual of $a$-maximization \cite{Intriligator:2003jj} developed in  \cite{Martelli:2005tp,Martelli:2006yb}). It would be interesting to develop an extension that applies for surface and general defect SCFTs. In particular for defects that correspond to branes wrapping submanifolds $\cS$ of the internal manifold $\cM_{\rm int}$ in the holographic dual, the candidate functional will involve the embedding of $\cS \subset \cM_{\rm int}$. 

For product geometries, namely when $\cM_{\rm int}$ is trivially fibered over the AdS base, the extremization problem appears to be trivial. 
For example, for 4d $\cN=1$ SCFTs dual to type IIB string theory on $AdS_5\times {\rm SE}_5$ with a Sasaki-Einstein (SE) internal manifold, a class of half-BPS $\cN=(0,2)$ surface defects correspond to D3 branes wrapping $AdS_3\times S^1$, where the $S^1 \subset {\rm SE}_5$
is required to be a closed orbit of the Reeb vector field $\xi$ to preserve supersymmetry \cite{Koh:2008kt}. With a single probe D3-brane, one naturally expects the conformal $b$-anomaly of the dual surface defect to be proportional to the size $\ell$ of $S^1$.
The Reeb vector $\xi$ realizes ${\rm SE}_5$ as a principal $U(1)$ bundle over a K\"ahler-Einstein base manifold $\cZ$,
and the orbits are simply the $S^1$ fibres labelled by points on $\cZ$.
Since the Reeb fibres are geodesics and have the same length $\ell$ \cite{Martelli:2006yb,Sparks:2010sn}, the D3 brane  wrapping these fibres should give rise to a family of surface defects with identical $b$-anomalies, potentially related by marginal defect deformations.\footnote{Here for simplicity we have assumed that ${\rm SE}_5$ is a \textit{regular} Sasaki-Einstein manifold but the statements here hold with small modifications for \textit{quasi-regular} Sasaki-Einstein manifolds. We refer the readers to \cite{Martelli:2006yb,Sparks:2010sn} for background material on these concepts.} It would be interesting to investigate such an $\cN=(0,2)$ defect conformal manifold.

\subsubsection*{Defect anomalies from defect correlation functions}
Both 't Hooft anomalies and conformal anomalies are physical observables of a given CFT. Although they are often defined as contact term modifications of symmetry Ward identities, the anomalies contribute unambiguously to the correlation functions of local (and sometimes extended) operators at separate points. For anomalies in the absence of defects, the connection between their contact-term and separated-point-correlation manifestations  is well established, while for defect anomalies, this is yet to be fully developed. 

For surface defects, it is straightforward to see that the two-point function of bulk conserved current $J_\m$ for a $U(1)_J$ symmetry has a unique parity-odd structure that is possible in $d=3$ with the defect along $y=0$,\footnote{Here we work with the Euclidean CFT obtained from Wick rotation, and thus the extra $i$ factor.}
\ie
\la J_\m(x) J_\n(x')\ra_{\cD}^{\rm odd} = { f(v)\over s^4} i\ep_{\m \rho \sigma }X^\rho I^\sigma{}_\n \,,
\fe
where $s\equiv x-x'$ and
\ie
v^2\equiv {(x-x')^2\over (x-x')^2+4yy'} 
\fe
is the invariant cross ratio under the residual conformal symmetry. The details on the tensor structure can be found in Appendix~\ref{app:podd}.
Since the defect OPE limit corresponds to $v\to 1$, we naturally expect $f(1)$ to encode the boundary $U(1)_J$ anomaly $k_{JJ}$. A simple computation of the two-point function for a 3d Dirac fermion and comparison to Table~\ref{tab:bta}, leads to the following conjecture
\ie
f(1)={1\over 2\pi^2} k_{JJ}\,.
\fe
A similar exercise can be done for the parity-odd contribution to the stress-tensor two-point function which should relate to the boundary gravitational anomaly.
These parity-odd defect two-point functions have also been studied in special kinematic regime in the momentum space \cite{Prochazka:2019bhv}, where connections to bulk Chern-Simons contact terms \cite{Closset:2012ru,Closset:2012vg} and defect 't Hooft anomalies were made. It would be desirable to compare with the position space approach here.

As for the surface defect $b$-anomaly, it was conjectured to be determined by the parity-even part of the stress-tensor two-point function $\la T_{\m\n}(x) T_{\A\B} (x')\ra_{\cD}$  in the defect OPE limit $v\to 1$ \cite{Herzog:2017kkj}. However this conjecture appears to be in tension with  large $N$ results for the Dirichlet boundary of the 3d $O(N)$ model as explained in the recent work \cite{Herzog:2020lel}.

 
\subsubsection*{Defects of dimensions $p\neq 2$}
Finally it will be interesting to investigate  anomalies and constraints on RG flows for defects of other longitudinal dimensions, namely $p\neq 2$. In particular, the $a$-theorem of \cite{Komargodski:2011vj,Komargodski:2011xv} has an immediate generalization to $p=4$  defect RG flows which will appear in \cite{bathm}. The $a$-maximization principle of \cite{Intriligator:2003jj} also has a natural extension to the superconformal defect similar to what we studied here \cite{bathm}. The story in odd defect dimensions is qualitatively different due to the absence of such $a$-type conformal anomalies. Instead it has been conjectured that the defect free energy $F_\cD$ plays the role of the monotonic function under RG flows, which have passed a number of tests (see \cite{Kobayashi:2018lil} for a recent summary). For half-BPS superconformal boundaries (interfaces) in 4d $\cN=2$ SCFTs, there is also a proposal of a boundary version \cite{Gaiotto:2014gha} of the $F$-maximization principle \cite{Jafferis:2010un} for 3d $\cN=2$ SCFTs. The proofs for these conjectures remain largely open in $d>2$ CFTs.


\section*{Acknowledgements}
The author thanks Nathan Agmon for collaboration on related topics. The author also thanks Zohar Komargodski for interesting comments on the draft.
The work of YW is  supported in part by the Center for Mathematical Sciences and Applications and the Center for the Fundamental Laws of Nature at Harvard University.

\appendix

\section{Boundary Feynman diagram}
\label{appendix:bdyf}
We would like to evaluate the integral
\ie
I(p,m) \equiv &\int {d^2 k \over (2\pi)^4}
{2\over (|k|+m) (|k+p|+m)}\,.
\fe
We proceed by introducing the Schwinger parameters $s_1,s_2$ and rewrite the integal as
\ie
I(p,m)=&
\int {d^2k \over (2\pi)^2} \int_0^\infty   ds_1 ds_2 \,
e^{-(s_1 |k|+s_2|k+p|) -m(s_1+s_2)}\,.
\fe
Next using the Laplace transform,
\ie
e^{-s|k|}=\int_0^\infty {dt\over 2\sqrt{\pi}}e^{-t k^2}   {s e^{-{s^2\over 4t}}\over t^{3/2}}\,,
\fe
we have
\ie
I(p,m)=&
{1\over 4\pi}\int {d^2k \over (2\pi)^2} \int_0^\infty  d s_1 ds_2 \, s_1s_2 e^{-m(s_1+s_2)}}\int_0^\infty  {dt_1dt_2\over (t_1t_2)^{3\over 2}}
e^{-t_1  k^2-t_2|k+p|^2} {e^{-{s_1^2\over 4t_1}-{s_2^2\over 4t_2}}\,.
\fe
Performing the $k$ integral and rescaling the variables $s_i \to s_i/m,t_i\to t_i/m^2$, we have
\ie
I(p,m)
 =&
{1\over (4\pi)^2} \int_0^\infty  d s_1 ds_2 \, s_1s_2 e^{-(s_1+s_2)}}
\int_0^\infty  {dt_1dt_2\over (t_1t_2)^{3\over 2}}
{e^{-{s_1^2\over 4t_1}-{s_2^2\over 4t_2}}
 {1\over t_1+t_2}e^{-{t_1 t_2 \over t_1+t_2} {p^2\over m^2}  }\,.
\fe
Now for the $p^2\over m^2$ term has coefficient after a change of variables $t_i \to 1/t_i$,
\ie
\A_2= 
 &
-{2\over (4\pi)^2} \int_0^\infty  d s_1 ds_2 \, s_1s_2 e^{-(s_1+s_2)}}
 \int_0^\infty  {dt_1dt_2}
 {e^{-{s_1^2\over 4}t_1-{s_2^2\over 4}t_1}
  { (t_1t_2)^{1\over 2}\over (t_1+t_2)^2} \,.
 \fe
 We further make the change of variables $t=t_1+t_2$ and $u={t_1\over t_1+t_2}$,
 \ie
 \A_2=&
-{2\over (4\pi)^2}  \int_0^\infty  ds_1ds_2  e^{- (s_1+s_2)}s_1 s_2\int_0^\infty dt \int_0^1 du  
  e^{-{1\over 4}t(s_1^2 u +s_2^2 (1-u))} \sqrt{u(1-u)}
  \\
   =&
-{1\over 4\pi}  \int_0^\infty  ds_1ds_2  e^{- (s_1+s_2)}{s_1 s_2\over (s_1+s_2)^2}\,.
\fe 
where we first do the $t$ integral followed by the $u$ integral. The leftover integral can be done easily again by a change of variables similar to what we have used for $t_i$, yielding
\ie
\A_2=-{1\over 4\pi} \int_0^\infty ds \int_0^1 du \,s e^{-s} u(1-u) =-{1\over 24 \pi}\,.
\fe

\section{Parity-violating boundary two-point function}
\label{app:podd}

The two-point function of bulk conserved currents in a general $d$-dimensional Euclidean CFT with a conformal boundary condition $\cD$ has a unique parity-even structure \cite{McAvity:1995zd,Herzog:2017xha}
\ie
\la J_\m (x) J_\n (x')\ra_\cD^{\rm even} ={1\over s^{2(d-1)}}\left(\pi(v) I_{\m\n}(s)-{1 \over d-1}v \pa_v \pi(v) \hat I_{\m\n} (s)\right )\,,
\label{JJeven}
\fe
where $s\equiv x-x'$ and 
\ie
v^2\equiv {(x-x')^2\over (x-x')^2+4yy'}
\fe
or equivalently
\ie 
 \xi\equiv {(x-x')^2\over 4y y'}~{\rm with}~ v^2={\xi\over \xi+1}
\fe
define the invariant cross-ratio under the residual $O(d,1)$ conformal symmetry. The tensor structure in \eqref{JJeven} involves the familiar Jacobian factor of inversion 
\ie
I_{\m\n}(x)=\D_{\m\n}-2 {x_\m x_\n \over x^2}\,,
\fe
and its modification due to the boundary defect,
\ie
\hat I_{\m\n}(x)\equiv \D_{\m\n}-X_\m X'_n,
\quad X_\m\equiv v\left ({2y\over s^2} s_\m - \D_{\m 3}  \right),\quad X'_\m \equiv v\left (-{2y'\over s^2} s_\m - \D_{\m 3}  \right)\,,
\fe
which transform nicely under bilocal $O(d,1)$ actions on $x$ and $x'$ \cite{McAvity:1995zd}.
Finally $\pi(v)$ is a general function of the cross-ratio $v$ subject to regularity constraints in the bulk OPE limit $v\to 0$ and the boundary OPE limit $v\to 1$ \cite{Herzog:2017xha}.

For $d=3$, another tensor structure becomes possible that uses the $\ep_{\m\n\rho}$ tensor
\ie
\la J_\m (x) J_\n (x')\ra_\cD^{\rm odd}
= { f(v)\over s^4} i\ep_{\m \rho \sigma }X^\rho I^\sigma{}_\n (s)\,,
\label{JJodd}
\fe
which has the right $O(d,1)$ transformation properties and satisfies current conservation. 
Note that another similar tensor structure built out of $I_{\m\n},X_\m,X'_\n$ is not independent
\ie
\ep_{\m \rho \sigma }X^\rho I^\sigma{}_\n(s) =\ep_{\n}{}^{\rho \sigma }X'_\rho I_{\sigma\m}\,. 
\fe
As discussed in the main text, we expect the function $f(v)$ in the defect OPE limit $v\to 1$ to encode the boundary 't Hooft anomaly of the $U(1)_J$ symmetry,
\ie
f(1) = \A k_{JJ}\,
\label{fkrel}
\fe
for some theory independent constant $\A$. 

To fix $\A$, let us consider the free Dirac fermion $\psi$ on $\mR^{2,1}_+$ with conformal boundary conditions $\left. \psi_+ \right |_\Sigma=0$ or $\left. \psi_- \right |_\Sigma=0$ which we refer to as $\cD_+$ and $\cD_-$ respectively. Here the Euclidean gamma matrices are chosen to be
\ie
\C=(\sigma_1,\sigma_2,\sigma_3)
\fe
and $\pm$ labels the eigenvalues of $\C_3$ which coincides with the chirality on the 2d boundary. The free fermion theory has a $U(1)_J$ global symmetry generated by
\ie
J_\m=\bar \psi \C_\m \psi\,,
\fe
under which $\psi$ have charge $+1$. The $U(1)_J$ symmetry is clearly preserved by the $\cD_\pm$ boundary conditions.  

According to general discussions in \cite{Dimofte:2017tpi} (see Table~\ref{tab:bta}), the $\cD_\pm$ boundary conditions contribute a boundary 't Hooft anomaly which is detectable in the presence of nontrivial $U(1)_J$ background gauge field,
\ie
k_{JJ}[\cD_\pm] = \mp {1\over 2}\,. 
\label{kJJD}
\fe

Now we compute $\la J_\m (x) J_\n (x')\ra_{\cD_\pm}$ explicitly using the free fermion propagator \cite{McAvity:1993ue}
\ie
\la \psi(x) \bar\psi(x')\ra_{\cD_\pm} ={i\over 4\pi} \left (
{\gamma \cdot (x-x')\over |x-x'|^3}
\pm
\C_3
{\gamma \cdot (\bar x-x')\over |\bar x-x'|^3}
\right)
\fe
where $\bar x\equiv (x,-y)$ is the reflection of $x$ across the boundary at $y=0$.

Performing the Wick contraction, we obtain after some algebra
\ie
\la J_\m (x) J_\n (x')\ra_{\cD_\pm }
=&-{1\over (4\pi)^2}
{1\over s^4}(-2(1+v^4)I_{\m\n} +4 v^4 X_\m X'_\n \pm 4 i v^2 \ep_{\m \A \B}X^\A I^\B{}_\n )
\\
=&
{1\over 8\pi^2 s^4}((1-v^4)I_{\m\n} +2 v^4 \hat I_{\m\n} \mp 2 i v^2 \ep_{\m \A \B}X^\A I^\B{}_\n )\,.
\fe
 Compared to \eqref{JJeven} and \eqref{JJodd}, we find for the boundary conditions $\cD_\pm$,\footnote{Note  $\pi(v)$ computed for free Dirac fermion in \cite{Herzog:2017xha} contains a a typo. For Dirac fermion in $d$-dimensions, $\pi(v) \propto 1-v^{2(d-1)}$ up to a constant.}
\ie
\pi(v)={1\over 8\pi^2}(1-v^4),\quad f_\pm (v)= \mp {1\over  4\pi^2}v^2\,.
\fe
Using \eqref{kJJD}, we thus conclude in \eqref{fkrel},
\ie
\A={1\over 2\pi^2}\,.
\fe

\bibliographystyle{JHEP}
\bibliography{defREF,SYMdefect}

\providecommand{\href}[2]{#2}\begingroup\raggedright\begin{thebibliography}{100}

\bibitem{tHooft:1979rat}
G.~'t~Hooft, \emph{{Naturalness, chiral symmetry, and spontaneous chiral
  symmetry breaking}},
  \href{https://doi.org/10.1007/978-1-4684-7571-5_9}{\emph{NATO Sci. Ser. B}
  {\bfseries 59} (1980) 135}.

\bibitem{Wess:1971yu}
J.~Wess and B.~Zumino, \emph{{Consequences of anomalous Ward identities}},
  \href{https://doi.org/10.1016/0370-2693(71)90582-X}{\emph{Phys. Lett. B}
  {\bfseries 37} (1971) 95}.

\bibitem{Zamolodchikov:1986gt}
A.~Zamolodchikov, \emph{{Irreversibility of the Flux of the Renormalization
  Group in a 2D Field Theory}}, {\emph{JETP Lett.} {\bfseries 43} (1986) 730}.

\bibitem{Cardy:1988cwa}
J.~L. Cardy, \emph{{Is There a c Theorem in Four-Dimensions?}},
  \href{https://doi.org/10.1016/0370-2693(88)90054-8}{\emph{Phys. Lett. B}
  {\bfseries 215} (1988) 749}.

\bibitem{Komargodski:2011vj}
Z.~Komargodski and A.~Schwimmer, \emph{{On Renormalization Group Flows in Four
  Dimensions}}, \href{https://doi.org/10.1007/JHEP12(2011)099}{\emph{JHEP}
  {\bfseries 12} (2011) 099} [\href{https://arxiv.org/abs/1107.3987}{{\ttfamily
  1107.3987}}].

\bibitem{Komargodski:2011xv}
Z.~Komargodski, \emph{{The Constraints of Conformal Symmetry on RG Flows}},
  \href{https://doi.org/10.1007/JHEP07(2012)069}{\emph{JHEP} {\bfseries 07}
  (2012) 069} [\href{https://arxiv.org/abs/1112.4538}{{\ttfamily 1112.4538}}].

\bibitem{Casini:2017vbe}
H.~Casini, E.~Test\'e and G.~Torroba, \emph{{Markov Property of the Conformal
  Field Theory Vacuum and the a Theorem}},
  \href{https://doi.org/10.1103/PhysRevLett.118.261602}{\emph{Phys. Rev. Lett.}
  {\bfseries 118} (2017) 261602}
  [\href{https://arxiv.org/abs/1704.01870}{{\ttfamily 1704.01870}}].

\bibitem{Elvang:2012st}
H.~Elvang, D.~Z. Freedman, L.-Y. Hung, M.~Kiermaier, R.~C. Myers and
  S.~Theisen, \emph{{On renormalization group flows and the a-theorem in 6d}},
  \href{https://doi.org/10.1007/JHEP10(2012)011}{\emph{JHEP} {\bfseries 10}
  (2012) 011} [\href{https://arxiv.org/abs/1205.3994}{{\ttfamily 1205.3994}}].

\bibitem{Cordova:2015vwa}
C.~Cordova, T.~T. Dumitrescu and X.~Yin, \emph{{Higher derivative terms,
  toroidal compactification, and Weyl anomalies in six-dimensional (2, 0)
  theories}}, \href{https://doi.org/10.1007/JHEP10(2019)128}{\emph{JHEP}
  {\bfseries 10} (2019) 128}
  [\href{https://arxiv.org/abs/1505.03850}{{\ttfamily 1505.03850}}].

\bibitem{Cordova:2015fha}
C.~Cordova, T.~T. Dumitrescu and K.~Intriligator, \emph{{Anomalies,
  renormalization group flows, and the a-theorem in six-dimensional (1, 0)
  theories}}, \href{https://doi.org/10.1007/JHEP10(2016)080}{\emph{JHEP}
  {\bfseries 10} (2016) 080}
  [\href{https://arxiv.org/abs/1506.03807}{{\ttfamily 1506.03807}}].

\bibitem{Graham:1999pm}
C.~Graham and E.~Witten, \emph{{Conformal anomaly of submanifold observables in
  AdS / CFT correspondence}},
  \href{https://doi.org/10.1016/S0550-3213(99)00055-3}{\emph{Nucl. Phys. B}
  {\bfseries 546} (1999) 52}
  [\href{https://arxiv.org/abs/hep-th/9901021}{{\ttfamily hep-th/9901021}}].

\bibitem{Henningson:1999xi}
M.~Henningson and K.~Skenderis, \emph{{Weyl anomaly for Wilson surfaces}},
  \href{https://doi.org/10.1088/1126-6708/1999/06/012}{\emph{JHEP} {\bfseries
  06} (1999) 012} [\href{https://arxiv.org/abs/hep-th/9905163}{{\ttfamily
  hep-th/9905163}}].

\bibitem{Schwimmer:2008yh}
A.~Schwimmer and S.~Theisen, \emph{{Entanglement Entropy, Trace Anomalies and
  Holography}},
  \href{https://doi.org/10.1016/j.nuclphysb.2008.04.015}{\emph{Nucl. Phys. B}
  {\bfseries 801} (2008) 1} [\href{https://arxiv.org/abs/0802.1017}{{\ttfamily
  0802.1017}}].

\bibitem{Jensen:2015swa}
K.~Jensen and A.~O'Bannon, \emph{{Constraint on Defect and Boundary
  Renormalization Group Flows}},
  \href{https://doi.org/10.1103/PhysRevLett.116.091601}{\emph{Phys. Rev. Lett.}
  {\bfseries 116} (2016) 091601}
  [\href{https://arxiv.org/abs/1509.02160}{{\ttfamily 1509.02160}}].

\bibitem{Vassilevich:2003xt}
D.~Vassilevich, \emph{{Heat kernel expansion: User's manual}},
  \href{https://doi.org/10.1016/j.physrep.2003.09.002}{\emph{Phys. Rept.}
  {\bfseries 388} (2003) 279}
  [\href{https://arxiv.org/abs/hep-th/0306138}{{\ttfamily hep-th/0306138}}].

\bibitem{Benini:2012cz}
F.~Benini and N.~Bobev, \emph{{Exact two-dimensional superconformal R-symmetry
  and c-extremization}},
  \href{https://doi.org/10.1103/PhysRevLett.110.061601}{\emph{Phys. Rev. Lett.}
  {\bfseries 110} (2013) 061601}
  [\href{https://arxiv.org/abs/1211.4030}{{\ttfamily 1211.4030}}].

\bibitem{Benini:2013cda}
F.~Benini and N.~Bobev, \emph{{Two-dimensional SCFTs from wrapped branes and
  c-extremization}}, \href{https://doi.org/10.1007/JHEP06(2013)005}{\emph{JHEP}
  {\bfseries 06} (2013) 005} [\href{https://arxiv.org/abs/1302.4451}{{\ttfamily
  1302.4451}}].

\bibitem{Gentle:2015jma}
S.~A. Gentle, M.~Gutperle and C.~Marasinou, \emph{{Entanglement entropy of
  Wilson surfaces from bubbling geometries in M-theory}},
  \href{https://doi.org/10.1007/JHEP08(2015)019}{\emph{JHEP} {\bfseries 08}
  (2015) 019} [\href{https://arxiv.org/abs/1506.00052}{{\ttfamily
  1506.00052}}].

\bibitem{Rodgers:2018mvq}
R.~Rodgers, \emph{{Holographic entanglement entropy from probe M-theory
  branes}}, \href{https://doi.org/10.1007/JHEP03(2019)092}{\emph{JHEP}
  {\bfseries 03} (2019) 092}
  [\href{https://arxiv.org/abs/1811.12375}{{\ttfamily 1811.12375}}].

\bibitem{Jensen:2018rxu}
K.~Jensen, A.~O'Bannon, B.~Robinson and R.~Rodgers, \emph{{From the Weyl
  Anomaly to Entropy of Two-Dimensional Boundaries and Defects}},
  \href{https://doi.org/10.1103/PhysRevLett.122.241602}{\emph{Phys. Rev. Lett.}
  {\bfseries 122} (2019) 241602}
  [\href{https://arxiv.org/abs/1812.08745}{{\ttfamily 1812.08745}}].

\bibitem{Estes:2018tnu}
J.~Estes, D.~Krym, A.~O'Bannon, B.~Robinson and R.~Rodgers, \emph{{Wilson
  Surface Central Charge from Holographic Entanglement Entropy}},
  \href{https://doi.org/10.1007/JHEP05(2019)032}{\emph{JHEP} {\bfseries 05}
  (2019) 032} [\href{https://arxiv.org/abs/1812.00923}{{\ttfamily
  1812.00923}}].

\bibitem{Chalabi:2020iie}
A.~Chalabi, A.~O'Bannon, B.~Robinson and J.~Sisti, \emph{{Central charges of 2d
  superconformal defects}},
  \href{https://doi.org/10.1007/JHEP05(2020)095}{\emph{JHEP} {\bfseries 05}
  (2020) 095} [\href{https://arxiv.org/abs/2003.02857}{{\ttfamily
  2003.02857}}].

\bibitem{nw2}
N.~B. Agmon and Y.~Wang, \emph{{Classifying Superconformal Defects in Diverse
  Dimensions Part II: Superconformal Defects of Dimension $p>1$}},
  \href{https://arxiv.org/abs/to appear}{{\ttfamily to appear}}.

\bibitem{Bardeen:1984pm}
W.~A. Bardeen and B.~Zumino, \emph{{Consistent and Covariant Anomalies in Gauge
  and Gravitational Theories}},
  \href{https://doi.org/10.1016/0550-3213(84)90322-5}{\emph{Nucl. Phys. B}
  {\bfseries 244} (1984) 421}.

\bibitem{Gomis:2015yaa}
J.~Gomis, P.-S. Hsin, Z.~Komargodski, A.~Schwimmer, N.~Seiberg and S.~Theisen,
  \emph{{Anomalies, Conformal Manifolds, and Spheres}},
  \href{https://doi.org/10.1007/JHEP03(2016)022}{\emph{JHEP} {\bfseries 03}
  (2016) 022} [\href{https://arxiv.org/abs/1509.08511}{{\ttfamily
  1509.08511}}].

\bibitem{Rocek:1985bk}
M.~Rocek and P.~van Nieuwenhuizen, \emph{{$N\geq 2$ supersymmetric Chern-Simons
  terms as d=3 extended conformal supergravity}},
  \href{https://doi.org/10.1088/0264-9381/3/1/007}{\emph{Classical and Quantum
  Gravity} {\bfseries 3} (1986) 43}.

\bibitem{Achucarro:1987vz}
A.~Achucarro and P.~Townsend, \emph{{A Chern-Simons Action for
  Three-Dimensional anti-De Sitter Supergravity Theories}},
  \href{https://doi.org/10.1016/0370-2693(86)90140-1}{\emph{Phys. Lett. B}
  {\bfseries 180} (1986) 89}.

\bibitem{Achucarro:1989gm}
A.~Achucarro and P.~Townsend, \emph{{Extended Supergravities in $d$ = (2+1) as
  \{Chern-Simons\} Theories}},
  \href{https://doi.org/10.1016/0370-2693(89)90423-1}{\emph{Phys. Lett. B}
  {\bfseries 229} (1989) 383}.

\bibitem{Closset:2012ru}
C.~Closset, T.~T. Dumitrescu, G.~Festuccia and Z.~Komargodski,
  \emph{{Supersymmetric Field Theories on Three-Manifolds}},
  \href{https://doi.org/10.1007/JHEP05(2013)017}{\emph{JHEP} {\bfseries 1305}
  (2013) 017} [\href{https://arxiv.org/abs/1212.3388}{{\ttfamily 1212.3388}}].

\bibitem{Closset:2012vg}
C.~Closset, T.~T. Dumitrescu, G.~Festuccia, Z.~Komargodski and N.~Seiberg,
  \emph{{Contact Terms, Unitarity, and F-Maximization in Three-Dimensional
  Superconformal Theories}},
  \href{https://doi.org/10.1007/JHEP10(2012)053}{\emph{JHEP} {\bfseries 1210}
  (2012) 053} [\href{https://arxiv.org/abs/1205.4142}{{\ttfamily 1205.4142}}].

\bibitem{Kraus:2005zm}
P.~Kraus and F.~Larsen, \emph{{Holographic gravitational anomalies}},
  \href{https://doi.org/10.1088/1126-6708/2006/01/022}{\emph{JHEP} {\bfseries
  01} (2006) 022} [\href{https://arxiv.org/abs/hep-th/0508218}{{\ttfamily
  hep-th/0508218}}].

\bibitem{Kraus:2006wn}
P.~Kraus, \emph{{Lectures on black holes and the AdS(3) / CFT(2)
  correspondence}}, {\emph{Lect. Notes Phys.} {\bfseries 755} (2008) 193}
  [\href{https://arxiv.org/abs/hep-th/0609074}{{\ttfamily hep-th/0609074}}].

\bibitem{Belyaev:2007bg}
D.~V. Belyaev and P.~van Nieuwenhuizen, \emph{{Tensor calculus for supergravity
  on a manifold with boundary}},
  \href{https://doi.org/10.1088/1126-6708/2008/02/047}{\emph{JHEP} {\bfseries
  02} (2008) 047} [\href{https://arxiv.org/abs/0711.2272}{{\ttfamily
  0711.2272}}].

\bibitem{Chamseddine:1992ry}
A.~H. Chamseddine and J.~Frohlich, \emph{{Two-dimensional Lorentz-Weyl anomaly
  and gravitational Chern-Simons theory}},
  \href{https://doi.org/10.1007/BF02097242}{\emph{Commun. Math. Phys.}
  {\bfseries 147} (1992) 549}.

\bibitem{Drukker:2017dgn}
N.~Drukker, I.~Shamir and C.~Vergu, \emph{{Defect multiplets of $ \mathcal{N}=1
  $ supersymmetry in 4d}},
  \href{https://doi.org/10.1007/JHEP01(2018)034}{\emph{JHEP} {\bfseries 01}
  (2018) 034} [\href{https://arxiv.org/abs/1711.03455}{{\ttfamily
  1711.03455}}].

\bibitem{Brunner:2019qyf}
I.~Brunner, J.~Schulz and A.~Tabler, \emph{{Boundaries and supercurrent
  multiplets in 3D Landau-Ginzburg models}},
  \href{https://doi.org/10.1007/JHEP06(2019)046}{\emph{JHEP} {\bfseries 06}
  (2019) 046} [\href{https://arxiv.org/abs/1904.07258}{{\ttfamily
  1904.07258}}].

\bibitem{Billo:2016cpy}
M.~Bill{\`o}, V.~Gon{\c c}alves, E.~Lauria and M.~Meineri, \emph{{Defects in
  conformal field theory}},
  \href{https://doi.org/10.1007/JHEP04(2016)091}{\emph{JHEP} {\bfseries 04}
  (2016) 091} [\href{https://arxiv.org/abs/1601.02883}{{\ttfamily
  1601.02883}}].

\bibitem{Herzog:2020bqw}
C.~P. Herzog and A.~Shrestha, \emph{{Two Point Functions in Defect CFTs}},
  \href{https://arxiv.org/abs/2010.04995}{{\ttfamily 2010.04995}}.

\bibitem{Dumitrescu:2011iu}
T.~T. Dumitrescu and N.~Seiberg, \emph{{Supercurrents and Brane Currents in
  Diverse Dimensions}},
  \href{https://doi.org/10.1007/JHEP07(2011)095}{\emph{JHEP} {\bfseries 07}
  (2011) 095} [\href{https://arxiv.org/abs/1106.0031}{{\ttfamily 1106.0031}}].

\bibitem{Kutasov:1990sv}
D.~Kutasov and N.~Seiberg, \emph{{Number of degrees of freedom, density of
  states and tachyons in string theory and CFT}},
  \href{https://doi.org/10.1016/0550-3213(91)90426-X}{\emph{Nucl. Phys. B}
  {\bfseries 358} (1991) 600}.

\bibitem{Hori:2001ax}
K.~Hori and A.~Kapustin, \emph{{Duality of the fermionic 2-D black hole and N=2
  liouville theory as mirror symmetry}},
  \href{https://doi.org/10.1088/1126-6708/2001/08/045}{\emph{JHEP} {\bfseries
  08} (2001) 045} [\href{https://arxiv.org/abs/hep-th/0104202}{{\ttfamily
  hep-th/0104202}}].

\bibitem{Adams:2001sv}
A.~Adams, J.~Polchinski and E.~Silverstein, \emph{{Don't panic! Closed string
  tachyons in ALE space-times}},
  \href{https://doi.org/10.1088/1126-6708/2001/10/029}{\emph{JHEP} {\bfseries
  10} (2001) 029} [\href{https://arxiv.org/abs/hep-th/0108075}{{\ttfamily
  hep-th/0108075}}].

\bibitem{Harvey:2001wm}
J.~A. Harvey, D.~Kutasov, E.~J. Martinec and G.~W. Moore, \emph{{Localized
  tachyons and RG flows}},
  \href{https://arxiv.org/abs/hep-th/0111154}{{\ttfamily hep-th/0111154}}.

\bibitem{Bianchi:2019umv}
L.~Bianchi, \emph{{Marginal deformations and defect anomalies}},
  \href{https://doi.org/10.1103/PhysRevD.100.126018}{\emph{Phys. Rev. D}
  {\bfseries 100} (2019) 126018}
  [\href{https://arxiv.org/abs/1907.06193}{{\ttfamily 1907.06193}}].

\bibitem{Bianchi:2019sxz}
L.~Bianchi and M.~Lemos, \emph{{Superconformal surfaces in four dimensions}},
  \href{https://doi.org/10.1007/JHEP06(2020)056}{\emph{JHEP} {\bfseries 06}
  (2020) 056} [\href{https://arxiv.org/abs/1911.05082}{{\ttfamily
  1911.05082}}].

\bibitem{Lewkowycz:2014jia}
A.~Lewkowycz and E.~Perlmutter, \emph{{Universality in the geometric dependence
  of Renyi entropy}},
  \href{https://doi.org/10.1007/JHEP01(2015)080}{\emph{JHEP} {\bfseries 01}
  (2015) 080} [\href{https://arxiv.org/abs/1407.8171}{{\ttfamily 1407.8171}}].

\bibitem{Bianchi:2015liz}
L.~Bianchi, M.~Meineri, R.~C. Myers and M.~Smolkin, \emph{{R\'enyi entropy and
  conformal defects}},
  \href{https://doi.org/10.1007/JHEP07(2016)076}{\emph{JHEP} {\bfseries 07}
  (2016) 076} [\href{https://arxiv.org/abs/1511.06713}{{\ttfamily
  1511.06713}}].

\bibitem{Beem:2013sza}
C.~Beem, M.~Lemos, P.~Liendo, W.~Peelaers, L.~Rastelli and B.~C. van Rees,
  \emph{{Infinite Chiral Symmetry in Four Dimensions}},
  \href{https://doi.org/10.1007/s00220-014-2272-x}{\emph{Commun. Math. Phys.}
  {\bfseries 336} (2015) 1359}
  [\href{https://arxiv.org/abs/1312.5344}{{\ttfamily 1312.5344}}].

\bibitem{Cordova:2017mhb}
C.~Cordova, D.~Gaiotto and S.-H. Shao, \emph{{Surface Defects and Chiral
  Algebras}}, \href{https://doi.org/10.1007/JHEP05(2017)140}{\emph{JHEP}
  {\bfseries 05} (2017) 140}
  [\href{https://arxiv.org/abs/1704.01955}{{\ttfamily 1704.01955}}].

\bibitem{Nekrasov:2003rj}
N.~Nekrasov and A.~Okounkov, \emph{{Seiberg-Witten theory and random
  partitions}}, \href{https://doi.org/10.1007/0-8176-4467-9_15}{\emph{Prog.
  Math.} {\bfseries 244} (2006) 525}
  [\href{https://arxiv.org/abs/hep-th/0306238}{{\ttfamily hep-th/0306238}}].

\bibitem{Pestun:2007rz}
V.~Pestun, \emph{{Localization of gauge theory on a four-sphere and
  supersymmetric Wilson loops}},
  \href{https://doi.org/10.1007/s00220-012-1485-0}{\emph{Commun. Math. Phys.}
  {\bfseries 313} (2012) 71} [\href{https://arxiv.org/abs/0712.2824}{{\ttfamily
  0712.2824}}].

\bibitem{Drukker:2010jp}
N.~Drukker, D.~Gaiotto and J.~Gomis, \emph{{The Virtue of Defects in 4D Gauge
  Theories and 2D CFTs}},
  \href{https://doi.org/10.1007/JHEP06(2011)025}{\emph{JHEP} {\bfseries 06}
  (2011) 025} [\href{https://arxiv.org/abs/1003.1112}{{\ttfamily 1003.1112}}].

\bibitem{Kanno:2011fw}
H.~Kanno and Y.~Tachikawa, \emph{{Instanton counting with a surface operator
  and the chain-saw quiver}},
  \href{https://doi.org/10.1007/JHEP06(2011)119}{\emph{JHEP} {\bfseries 06}
  (2011) 119} [\href{https://arxiv.org/abs/1105.0357}{{\ttfamily 1105.0357}}].

\bibitem{Gomis:2014eya}
J.~Gomis and B.~Le~Floch, \emph{{M2-brane surface operators and gauge theory
  dualities in Toda}},
  \href{https://doi.org/10.1007/JHEP04(2016)183}{\emph{JHEP} {\bfseries 04}
  (2016) 183} [\href{https://arxiv.org/abs/1407.1852}{{\ttfamily 1407.1852}}].

\bibitem{Gorsky:2017hro}
A.~Gorsky, B.~Le~Floch, A.~Milekhin and N.~Sopenko, \emph{{Surface defects and
  instanton\textendash{}vortex interaction}},
  \href{https://doi.org/10.1016/j.nuclphysb.2017.04.010}{\emph{Nucl. Phys. B}
  {\bfseries 920} (2017) 122}
  [\href{https://arxiv.org/abs/1702.03330}{{\ttfamily 1702.03330}}].

\bibitem{Berenstein:1998ij}
D.~E. Berenstein, R.~Corrado, W.~Fischler and J.~M. Maldacena, \emph{{The
  Operator product expansion for Wilson loops and surfaces in the large N
  limit}}, \href{https://doi.org/10.1103/PhysRevD.59.105023}{\emph{Phys. Rev.
  D} {\bfseries 59} (1999) 105023}
  [\href{https://arxiv.org/abs/hep-th/9809188}{{\ttfamily hep-th/9809188}}].

\bibitem{Henningson:1999iw}
M.~Henningson, \emph{{Surface observables and the Weyl anomaly}},  in
  \emph{{14th International Workshop on High-Energy Physics and Quantum Field
  Theory (QFTHEP 99)}}, pp.~384--386, 5, 1999,
  \href{https://arxiv.org/abs/hep-th/9908183}{{\ttfamily hep-th/9908183}}.

\bibitem{Gustavsson:2003hn}
A.~Gustavsson, \emph{{On the Weyl anomaly of Wilson surfaces}},
  \href{https://doi.org/10.1088/1126-6708/2003/12/059}{\emph{JHEP} {\bfseries
  12} (2003) 059} [\href{https://arxiv.org/abs/hep-th/0310037}{{\ttfamily
  hep-th/0310037}}].

\bibitem{Gustavsson:2004gj}
A.~Gustavsson, \emph{{Conformal anomaly of Wilson surface observables: A Field
  theoretical computation}},
  \href{https://doi.org/10.1088/1126-6708/2004/07/074}{\emph{JHEP} {\bfseries
  07} (2004) 074} [\href{https://arxiv.org/abs/hep-th/0404150}{{\ttfamily
  hep-th/0404150}}].

\bibitem{Mezei:2018url}
M.~Mezei, S.~S. Pufu and Y.~Wang, \emph{{Chern-Simons theory from M5-branes and
  calibrated M2-branes}},
  \href{https://doi.org/10.1007/JHEP08(2019)165}{\emph{JHEP} {\bfseries 08}
  (2019) 165} [\href{https://arxiv.org/abs/1812.07572}{{\ttfamily
  1812.07572}}].

\bibitem{Drukker:2020dcz}
N.~Drukker, M.~Probst and M.~Tr\'epanier, \emph{{Surface operators in the 6d N
  = (2, 0) theory}}, \href{https://doi.org/10.1088/1751-8121/aba1b7}{\emph{J.
  Phys. A} {\bfseries 53} (2020) 365401}
  [\href{https://arxiv.org/abs/2003.12372}{{\ttfamily 2003.12372}}].

\bibitem{Drukker:2020swu}
N.~Drukker, S.~Giombi, A.~A. Tseytlin and X.~Zhou, \emph{{Defect CFT in the 6d
  (2,0) theory from M2 brane dynamics in AdS$_7 \times$S$^4$}},
  \href{https://doi.org/10.1007/JHEP07(2020)101}{\emph{JHEP} {\bfseries 07}
  (2020) 101} [\href{https://arxiv.org/abs/2004.04562}{{\ttfamily
  2004.04562}}].

\bibitem{Drukker:2020atp}
N.~Drukker, M.~Probst and M.~Tr\'epanier, \emph{{Defect CFT techniques in the
  6d $\mathcal{N} = (2,0)$ theory}},
  \href{https://arxiv.org/abs/2009.10732}{{\ttfamily 2009.10732}}.

\bibitem{Bullimore:2014upa}
M.~Bullimore and H.-C. Kim, \emph{{The Superconformal Index of the (2,0) Theory
  with Defects}}, \href{https://doi.org/10.1007/JHEP05(2015)048}{\emph{JHEP}
  {\bfseries 05} (2015) 048} [\href{https://arxiv.org/abs/1412.3872}{{\ttfamily
  1412.3872}}].

\bibitem{Berman:2009kj}
D.~S. Berman and D.~C. Thompson, \emph{{Membranes with a boundary}},
  \href{https://doi.org/10.1016/j.nuclphysb.2009.06.004}{\emph{Nucl. Phys. B}
  {\bfseries 820} (2009) 503}
  [\href{https://arxiv.org/abs/0904.0241}{{\ttfamily 0904.0241}}].

\bibitem{Berman:2009xd}
D.~S. Berman, M.~J. Perry, E.~Sezgin and D.~C. Thompson, \emph{{Boundary
  Conditions for Interacting Membranes}},
  \href{https://doi.org/10.1007/JHEP04(2010)025}{\emph{JHEP} {\bfseries 04}
  (2010) 025} [\href{https://arxiv.org/abs/0912.3504}{{\ttfamily 0912.3504}}].

\bibitem{Dimofte:2012pd}
T.~Dimofte and D.~Gaiotto, \emph{{An E7 Surprise}},
  \href{https://doi.org/10.1007/JHEP10(2012)129}{\emph{JHEP} {\bfseries 10}
  (2012) 129} [\href{https://arxiv.org/abs/1209.1404}{{\ttfamily 1209.1404}}].

\bibitem{Gadde:2013wq}
A.~Gadde, S.~Gukov and P.~Putrov, \emph{{Walls, Lines, and Spectral Dualities
  in 3d Gauge Theories}},
  \href{https://doi.org/10.1007/JHEP05(2014)047}{\emph{JHEP} {\bfseries 05}
  (2014) 047} [\href{https://arxiv.org/abs/1302.0015}{{\ttfamily 1302.0015}}].

\bibitem{Gadde:2013sca}
A.~Gadde, S.~Gukov and P.~Putrov, \emph{{Fivebranes and 4-manifolds}},
  \href{https://doi.org/10.1007/978-3-319-43648-7_7}{\emph{Prog. Math.}
  {\bfseries 319} (2016) 155}
  [\href{https://arxiv.org/abs/1306.4320}{{\ttfamily 1306.4320}}].

\bibitem{Okazaki:2013kaa}
T.~Okazaki and S.~Yamaguchi, \emph{{Supersymmetric boundary conditions in
  three-dimensional N=2 theories}},
  \href{https://doi.org/10.1103/PhysRevD.87.125005}{\emph{Phys. Rev. D}
  {\bfseries 87} (2013) 125005}
  [\href{https://arxiv.org/abs/1302.6593}{{\ttfamily 1302.6593}}].

\bibitem{Yoshida:2014ssa}
Y.~Yoshida and K.~Sugiyama, \emph{{Localization of 3d $\mathcal{N}=2$
  Supersymmetric Theories on $S^1 \times D^2$}},
  \href{https://arxiv.org/abs/1409.6713}{{\ttfamily 1409.6713}}.

\bibitem{Bullimore:2016nji}
M.~Bullimore, T.~Dimofte, D.~Gaiotto and J.~Hilburn, \emph{{Boundaries, Mirror
  Symmetry, and Symplectic Duality in 3d $\mathcal{N}=4$ Gauge Theory}},
  \href{https://doi.org/10.1007/JHEP10(2016)108}{\emph{JHEP} {\bfseries 10}
  (2016) 108} [\href{https://arxiv.org/abs/1603.08382}{{\ttfamily
  1603.08382}}].

\bibitem{Dimofte:2017tpi}
T.~Dimofte, D.~Gaiotto and N.~M. Paquette, \emph{{Dual boundary conditions in
  3d SCFT\textquoteright{}s}},
  \href{https://doi.org/10.1007/JHEP05(2018)060}{\emph{JHEP} {\bfseries 05}
  (2018) 060} [\href{https://arxiv.org/abs/1712.07654}{{\ttfamily
  1712.07654}}].

\bibitem{Fursaev:2016inw}
D.~V. Fursaev and S.~N. Solodukhin, \emph{{Anomalies, entropy and boundaries}},
  \href{https://doi.org/10.1103/PhysRevD.93.084021}{\emph{Phys. Rev. D}
  {\bfseries 93} (2016) 084021}
  [\href{https://arxiv.org/abs/1601.06418}{{\ttfamily 1601.06418}}].

\bibitem{Seiberg:2016gmd}
N.~Seiberg, T.~Senthil, C.~Wang and E.~Witten, \emph{{A Duality Web in 2+1
  Dimensions and Condensed Matter Physics}},
  \href{https://doi.org/10.1016/j.aop.2016.08.007}{\emph{Annals Phys.}
  {\bfseries 374} (2016) 395}
  [\href{https://arxiv.org/abs/1606.01989}{{\ttfamily 1606.01989}}].

\bibitem{Koh:2009cj}
E.~Koh and S.~Yamaguchi, \emph{{Surface operators in the Klebanov-Witten
  theory}}, \href{https://doi.org/10.1088/1126-6708/2009/06/070}{\emph{JHEP}
  {\bfseries 06} (2009) 070} [\href{https://arxiv.org/abs/0904.1460}{{\ttfamily
  0904.1460}}].

\bibitem{Razamat:2018zel}
S.~S. Razamat, \emph{{Flavored surface defects in 4d $\mathcal{N}=1$ SCFTs}},
  \href{https://doi.org/10.1007/s11005-018-01145-9}{\emph{Lett. Math. Phys.}
  {\bfseries 109} (2019) 1377}
  [\href{https://arxiv.org/abs/1808.09509}{{\ttfamily 1808.09509}}].

\bibitem{Gukov:2006jk}
S.~Gukov and E.~Witten, \emph{{Gauge Theory, Ramification, And The Geometric
  Langlands Program}},  \href{https://arxiv.org/abs/hep-th/0612073}{{\ttfamily
  hep-th/0612073}}.

\bibitem{Gomis:2007fi}
J.~Gomis and S.~Matsuura, \emph{{Bubbling surface operators and S-duality}},
  \href{https://doi.org/10.1088/1126-6708/2007/06/025}{\emph{JHEP} {\bfseries
  06} (2007) 025} [\href{https://arxiv.org/abs/0704.1657}{{\ttfamily
  0704.1657}}].

\bibitem{Drukker:2008wr}
N.~Drukker, J.~Gomis and S.~Matsuura, \emph{{Probing N=4 SYM With Surface
  Operators}}, \href{https://doi.org/10.1088/1126-6708/2008/10/048}{\emph{JHEP}
  {\bfseries 10} (2008) 048} [\href{https://arxiv.org/abs/0805.4199}{{\ttfamily
  0805.4199}}].

\bibitem{Koh:2008kt}
E.~Koh and S.~Yamaguchi, \emph{{Holography of BPS surface operators}},
  \href{https://doi.org/10.1088/1126-6708/2009/02/012}{\emph{JHEP} {\bfseries
  02} (2009) 012} [\href{https://arxiv.org/abs/0812.1420}{{\ttfamily
  0812.1420}}].

\bibitem{Alday:2009fs}
L.~F. Alday, D.~Gaiotto, S.~Gukov, Y.~Tachikawa and H.~Verlinde, \emph{{Loop
  and surface operators in N=2 gauge theory and Liouville modular geometry}},
  \href{https://doi.org/10.1007/JHEP01(2010)113}{\emph{JHEP} {\bfseries 01}
  (2010) 113} [\href{https://arxiv.org/abs/0909.0945}{{\ttfamily 0909.0945}}].

\bibitem{Gaiotto:2009fs}
D.~Gaiotto, \emph{{Surface Operators in N = 2 4d Gauge Theories}},
  \href{https://doi.org/10.1007/JHEP11(2012)090}{\emph{JHEP} {\bfseries 11}
  (2012) 090} [\href{https://arxiv.org/abs/0911.1316}{{\ttfamily 0911.1316}}].

\bibitem{Gadde:2013dda}
A.~Gadde and S.~Gukov, \emph{{2d Index and Surface operators}},
  \href{https://doi.org/10.1007/JHEP03(2014)080}{\emph{JHEP} {\bfseries 03}
  (2014) 080} [\href{https://arxiv.org/abs/1305.0266}{{\ttfamily 1305.0266}}].

\bibitem{Witten:1997yu}
E.~Witten, \emph{{On the conformal field theory of the Higgs branch}},
  \href{https://doi.org/10.1088/1126-6708/1997/07/003}{\emph{JHEP} {\bfseries
  07} (1997) 003} [\href{https://arxiv.org/abs/hep-th/9707093}{{\ttfamily
  hep-th/9707093}}].

\bibitem{Lunin:2007ab}
O.~Lunin, \emph{{1/2-BPS states in M theory and defects in the dual CFTs}},
  \href{https://doi.org/10.1088/1126-6708/2007/10/014}{\emph{JHEP} {\bfseries
  10} (2007) 014} [\href{https://arxiv.org/abs/0704.3442}{{\ttfamily
  0704.3442}}].

\bibitem{DelZotto:2014hpa}
M.~Del~Zotto, J.~J. Heckman, A.~Tomasiello and C.~Vafa, \emph{{6d Conformal
  Matter}}, \href{https://doi.org/10.1007/JHEP02(2015)054}{\emph{JHEP}
  {\bfseries 02} (2015) 054} [\href{https://arxiv.org/abs/1407.6359}{{\ttfamily
  1407.6359}}].

\bibitem{Horava:1996ma}
P.~Horava and E.~Witten, \emph{{Eleven-dimensional supergravity on a manifold
  with boundary}},
  \href{https://doi.org/10.1016/0550-3213(96)00308-2}{\emph{Nucl. Phys. B}
  {\bfseries 475} (1996) 94}
  [\href{https://arxiv.org/abs/hep-th/9603142}{{\ttfamily hep-th/9603142}}].

\bibitem{Agmon:2020pde}
N.~B. Agmon and Y.~Wang, \emph{{Classifying Superconformal Defects in Diverse
  Dimensions Part I: Superconformal Lines}},
  \href{https://arxiv.org/abs/2009.06650}{{\ttfamily 2009.06650}}.

\bibitem{GREEN1985327}
M.~B. Green, J.~H. Schwarz and P.~West, \emph{Anomaly-free chiral theories in
  six dimensions},
  \href{https://doi.org/https://doi.org/10.1016/0550-3213(85)90222-6}{\emph{Nuclear
  Physics B} {\bfseries 254} (1985) 327 }.

\bibitem{Ohmori:2014kda}
K.~Ohmori, H.~Shimizu, Y.~Tachikawa and K.~Yonekura, \emph{{Anomaly polynomial
  of general 6d SCFTs}}, \href{https://doi.org/10.1093/ptep/ptu140}{\emph{PTEP}
  {\bfseries 2014} (2014) 103B07}
  [\href{https://arxiv.org/abs/1408.5572}{{\ttfamily 1408.5572}}].

\bibitem{Intriligator:2014eaa}
K.~Intriligator, \emph{{6d, $ \mathcal{N}=\left(1,\;0\right) $ Coulomb branch
  anomaly matching}},
  \href{https://doi.org/10.1007/JHEP10(2014)162}{\emph{JHEP} {\bfseries 10}
  (2014) 162} [\href{https://arxiv.org/abs/1408.6745}{{\ttfamily 1408.6745}}].

\bibitem{Haghighat:2013gba}
B.~Haghighat, A.~Iqbal, C.~Koz\c{c}az, G.~Lockhart and C.~Vafa,
  \emph{{M-Strings}},
  \href{https://doi.org/10.1007/s00220-014-2139-1}{\emph{Commun. Math. Phys.}
  {\bfseries 334} (2015) 779}
  [\href{https://arxiv.org/abs/1305.6322}{{\ttfamily 1305.6322}}].

\bibitem{Haghighat:2013tka}
B.~Haghighat, C.~Kozcaz, G.~Lockhart and C.~Vafa, \emph{{Orbifolds of
  M-strings}}, \href{https://doi.org/10.1103/PhysRevD.89.046003}{\emph{Phys.
  Rev. D} {\bfseries 89} (2014) 046003}
  [\href{https://arxiv.org/abs/1310.1185}{{\ttfamily 1310.1185}}].

\bibitem{Kim:2014dza}
J.~Kim, S.~Kim, K.~Lee, J.~Park and C.~Vafa, \emph{{Elliptic Genus of
  E-strings}}, \href{https://doi.org/10.1007/JHEP09(2017)098}{\emph{JHEP}
  {\bfseries 09} (2017) 098} [\href{https://arxiv.org/abs/1411.2324}{{\ttfamily
  1411.2324}}].

\bibitem{Haghighat:2014vxa}
B.~Haghighat, A.~Klemm, G.~Lockhart and C.~Vafa, \emph{{Strings of Minimal 6d
  SCFTs}}, \href{https://doi.org/10.1002/prop.201500014}{\emph{Fortsch. Phys.}
  {\bfseries 63} (2015) 294} [\href{https://arxiv.org/abs/1412.3152}{{\ttfamily
  1412.3152}}].

\bibitem{Gadde:2015tra}
A.~Gadde, B.~Haghighat, J.~Kim, S.~Kim, G.~Lockhart and C.~Vafa, \emph{{6d
  String Chains}}, \href{https://doi.org/10.1007/JHEP02(2018)143}{\emph{JHEP}
  {\bfseries 02} (2018) 143}
  [\href{https://arxiv.org/abs/1504.04614}{{\ttfamily 1504.04614}}].

\bibitem{Kim:2016foj}
H.-C. Kim, S.~Kim and J.~Park, \emph{{6d strings from new chiral gauge
  theories}},  \href{https://arxiv.org/abs/1608.03919}{{\ttfamily 1608.03919}}.

\bibitem{Shimizu:2016lbw}
H.~Shimizu and Y.~Tachikawa, \emph{{Anomaly of strings of 6d $
  \mathcal{N}=\left(1,0\right) $ theories}},
  \href{https://doi.org/10.1007/JHEP11(2016)165}{\emph{JHEP} {\bfseries 11}
  (2016) 165} [\href{https://arxiv.org/abs/1608.05894}{{\ttfamily
  1608.05894}}].

\bibitem{Couzens:2018wnk}
C.~Couzens, J.~P. Gauntlett, D.~Martelli and J.~Sparks, \emph{{A geometric dual
  of $c$-extremization}},
  \href{https://doi.org/10.1007/JHEP01(2019)212}{\emph{JHEP} {\bfseries 01}
  (2019) 212} [\href{https://arxiv.org/abs/1810.11026}{{\ttfamily
  1810.11026}}].

\bibitem{Gauntlett:2019roi}
J.~P. Gauntlett, D.~Martelli and J.~Sparks, \emph{{Toric geometry and the dual
  of ${\cal I}$-extremization}},
  \href{https://doi.org/10.1007/JHEP06(2019)140}{\emph{JHEP} {\bfseries 06}
  (2019) 140} [\href{https://arxiv.org/abs/1904.04282}{{\ttfamily
  1904.04282}}].

\bibitem{vanBeest:2020vlv}
M.~van Beest, S.~Cizel, S.~Schafer-Nameki and J.~Sparks,
  \emph{{$\mathcal{I}$/$c$-Extremization in M/F-Duality}},
  \href{https://doi.org/10.21468/SciPostPhys.9.3.029}{\emph{SciPost Phys.}
  {\bfseries 9} (2020) 029} [\href{https://arxiv.org/abs/2004.04020}{{\ttfamily
  2004.04020}}].

\bibitem{Intriligator:2003jj}
K.~A. Intriligator and B.~Wecht, \emph{{The Exact superconformal R symmetry
  maximizes a}},
  \href{https://doi.org/10.1016/S0550-3213(03)00459-0}{\emph{Nucl. Phys. B}
  {\bfseries 667} (2003) 183}
  [\href{https://arxiv.org/abs/hep-th/0304128}{{\ttfamily hep-th/0304128}}].

\bibitem{Martelli:2005tp}
D.~Martelli, J.~Sparks and S.-T. Yau, \emph{{The Geometric dual of
  a-maximisation for Toric Sasaki-Einstein manifolds}},
  \href{https://doi.org/10.1007/s00220-006-0087-0}{\emph{Commun. Math. Phys.}
  {\bfseries 268} (2006) 39}
  [\href{https://arxiv.org/abs/hep-th/0503183}{{\ttfamily hep-th/0503183}}].

\bibitem{Martelli:2006yb}
D.~Martelli, J.~Sparks and S.-T. Yau, \emph{{Sasaki-Einstein manifolds and
  volume minimisation}},
  \href{https://doi.org/10.1007/s00220-008-0479-4}{\emph{Commun. Math. Phys.}
  {\bfseries 280} (2008) 611}
  [\href{https://arxiv.org/abs/hep-th/0603021}{{\ttfamily hep-th/0603021}}].

\bibitem{Sparks:2010sn}
J.~Sparks, \emph{{Sasaki-Einstein Manifolds}},
  \href{https://doi.org/10.4310/SDG.2011.v16.n1.a6}{\emph{Surveys Diff. Geom.}
  {\bfseries 16} (2011) 265} [\href{https://arxiv.org/abs/1004.2461}{{\ttfamily
  1004.2461}}].

\bibitem{Prochazka:2019bhv}
V.~Prochazka, \emph{{Boundary gauge and gravitational anomalies from Ward
  identities}}, \href{https://doi.org/10.1007/JHEP07(2019)047}{\emph{JHEP}
  {\bfseries 07} (2019) 047}
  [\href{https://arxiv.org/abs/1901.10920}{{\ttfamily 1901.10920}}].

\bibitem{Herzog:2017kkj}
C.~Herzog, K.-W. Huang and K.~Jensen, \emph{{Displacement Operators and
  Constraints on Boundary Central Charges}},
  \href{https://doi.org/10.1103/PhysRevLett.120.021601}{\emph{Phys. Rev. Lett.}
  {\bfseries 120} (2018) 021601}
  [\href{https://arxiv.org/abs/1709.07431}{{\ttfamily 1709.07431}}].

\bibitem{Herzog:2020lel}
C.~P. Herzog and N.~Kobayashi, \emph{{The $O(N)$ model with $\phi^6$ potential
  in ${\mathbb R}^2 \times {\mathbb R}^+$}},
  \href{https://doi.org/10.1007/JHEP09(2020)126}{\emph{JHEP} {\bfseries 09}
  (2020) 126} [\href{https://arxiv.org/abs/2005.07863}{{\ttfamily
  2005.07863}}].

\bibitem{bathm}
Y.~Wang, \emph{{Defect $a$-theorem and $a$-maximization}},
  \href{https://arxiv.org/abs/to appear}{{\ttfamily to appear}}.

\bibitem{Kobayashi:2018lil}
N.~Kobayashi, T.~Nishioka, Y.~Sato and K.~Watanabe, \emph{{Towards a
  $C$-theorem in defect CFT}},
  \href{https://doi.org/10.1007/JHEP01(2019)039}{\emph{JHEP} {\bfseries 01}
  (2019) 039} [\href{https://arxiv.org/abs/1810.06995}{{\ttfamily
  1810.06995}}].

\bibitem{Gaiotto:2014gha}
D.~Gaiotto, \emph{{Boundary F-maximization}},
  \href{https://arxiv.org/abs/1403.8052}{{\ttfamily 1403.8052}}.

\bibitem{Jafferis:2010un}
D.~L. Jafferis, \emph{{The Exact Superconformal R-Symmetry Extremizes $Z$}},
  \href{https://doi.org/10.1007/JHEP05(2012)159}{\emph{JHEP} {\bfseries 1205}
  (2012) 159} [\href{https://arxiv.org/abs/1012.3210}{{\ttfamily 1012.3210}}].

\bibitem{McAvity:1995zd}
D.~McAvity and H.~Osborn, \emph{{Conformal field theories near a boundary in
  general dimensions}},
  \href{https://doi.org/10.1016/0550-3213(95)00476-9}{\emph{Nucl. Phys. B}
  {\bfseries 455} (1995) 522}
  [\href{https://arxiv.org/abs/cond-mat/9505127}{{\ttfamily
  cond-mat/9505127}}].

\bibitem{Herzog:2017xha}
C.~P. Herzog and K.-W. Huang, \emph{{Boundary Conformal Field Theory and a
  Boundary Central Charge}},
  \href{https://doi.org/10.1007/JHEP10(2017)189}{\emph{JHEP} {\bfseries 10}
  (2017) 189} [\href{https://arxiv.org/abs/1707.06224}{{\ttfamily
  1707.06224}}].

\bibitem{McAvity:1993ue}
D.~McAvity and H.~Osborn, \emph{{Energy momentum tensor in conformal field
  theories near a boundary}},
  \href{https://doi.org/10.1016/0550-3213(93)90005-A}{\emph{Nucl. Phys. B}
  {\bfseries 406} (1993) 655}
  [\href{https://arxiv.org/abs/hep-th/9302068}{{\ttfamily hep-th/9302068}}].

\end{thebibliography}\endgroup

\end{document}